\documentclass[12pt]{article} 

\usepackage{scicite}
\usepackage{times}

\usepackage{amsmath}
\usepackage{amssymb}
\usepackage{graphicx}
\usepackage{verbatim}
\usepackage[colorlinks=true, pdfstartview=FitV, linkcolor=blue, citecolor=blue, urlcolor=blue]{hyperref}

\newcommand{\ee}[1]{\cdot10^{#1}}
\newcommand{\mr}[1]{\mathrm{#1}}
\newcommand{\unit}[1]{\,\mathrm{#1}}
\newcommand{\um}{\,\mu{\rm m}}

\newcommand{\kB}{k_{\rm B}}
\newcommand{\mo}{\mu_0}

\newcommand{\acrit}{a_\mr{crit}}

\newcommand{\Bx}{B_x}
\newcommand{\By}{B_y}

\newcommand{\Dv}{D_\nu}
\newcommand{\EF}{E_\mr{F}}
\newcommand{\Isd}{I_0}

\newcommand{\JxTilde}{J'_x}

\newcommand{\JyTilde}{J'_y}

\newcommand{\vecJ}{\vec{J}}

\newcommand{\lb}{l_\mr{b}}
\newcommand{\lee}{l_\mr{ee}}
\newcommand{\lmr}{l_\mr{mr}}

\newcommand{\vecr}{\vec{r}}

\newenvironment{sciabstract}{%
\begin{quote} \bf}
{\end{quote}}

\title{Observation of current whirlpools in graphene \newline at room temperature} 

\author{
Marius~L.~Palm$^{1\dagger}$, Chaoxin~Ding$^{1\dagger}$, William~S.~Huxter$^{1\dagger}$, \\
Takashi Taniguchi$^{2}$, Kenji Watanabe$^{3}$, and Christian~L.~Degen$^{1,4\ast}$ \\
\\
\normalsize{$^1$Department of Physics, ETH Zurich, Otto Stern Weg 1, 8093 Zurich, Switzerland;} \\
\normalsize{$^2$Research Center for Materials Nanoarchitectonics, National Institute for Materials Science,} \\
  \normalsize{1-1 Namiki, Tsukuba 305-0044, Japan;} \\
\normalsize{$^3$Research Center for Electronic and Optical Materials, National Institute for Materials Science,} \\
  \normalsize{1-1 Namiki, Tsukuba 305-0044, Japan;} \\
\normalsize{$^4$Quantum Center, ETH Zurich, 8093 Zurich, Switzerland.} \\
\\
\normalsize{$^\ast$To whom correspondence should be addressed; E-mail: degenc@ethz.ch.} \\
\normalsize{$^\dagger$These authors contributed equally.}
}

\date{}

\begin{document}

\baselineskip24pt
\maketitle 

\begin{sciabstract}
Electron-electron interactions in high-mobility conductors can give rise to transport signatures resembling those described by classical hydrodynamics.  Using a nanoscale scanning magnetometer, we imaged a distinctive hydrodynamic transport pattern -- stationary current vortices -- in a monolayer graphene device at room temperature. By measuring devices with increasing characteristic size, we observed the disappearance of the current vortex and thus verify a prediction of the hydrodynamic model.  We further observed that vortex flow is present for both hole- and electron-dominated transport regimes, while disappearing in the ambipolar regime.  We attribute this effect to a reduction of the vorticity diffusion length near charge neutrality.  Our work showcases the power of local imaging techniques for unveiling exotic mesoscopic transport phenomena.
\end{sciabstract}
\clearpage

Transport phenomena in mesoscopic devices are governed by the relative distance separating carrier scattering events compared to the characteristic device size $L$.  In a non-interacting system, once the device size becomes smaller than the momentum-relaxing scattering length $\lmr$ set by collision events with impurities and phonons ($L\ll \lmr$), carriers move unimpeded until they are scattered off a device boundary. This ballistic regime is of great scientific interest and manifests itself, for example, in transverse magnetic focusing experiments~\cite{chen16science} or through a quantized conductance in quantum point contacts~\cite{vanwees88}. 

In contrast, momentum-conserving collisions between carriers play a minor role in the transport of conventional metals, because they occur much less frequently than momentum-relaxing collisions~\cite{lucas18,narozhny22}.   However, in materials where scattering events are scarce, such as encapsulated graphene and high-mobility Ga[Al]As heterostructures at intermediate temperatures, $\lmr$ can approach or even surpass the carrier-carrier scattering length ($\lee$) for a finite temperature range. Consequently, in a device satisfying $\lee \ll L,\lmr$, transport properties become dominated by carrier-carrier interactions.  This regime, governed by the collective behavior of interacting carriers, can give rise to peculiar transport features that are not expected when compared to traditional diffusive or ballistic transport, such as viscosity~\cite{levitov16} or even turbulence~\cite{disante20}.  Given its similarity to classical fluid flow, this transport regime is commonly referred to as the viscous or hydrodynamic regime.

Initial theoretical work on hydrodynamic electron transport predicted a decrease of the resistivity with increasing temperature in metallic wires~\cite{gurzhi68}. This effect, known as the Gurzhi effect, was first demonstrated experimentally in a Ga[Al]As heterostructure~\cite{molenkamp94,dejong95}.  Other hallmarks of hydrodynamic transport include the viscous Hall effect~\cite{berdyugin19, kim20, scaffidi17, pellegrino17}, superballistic conduction~\cite{guo17, krishnakumar17, ginzburg21}, flow without the Landauer-Sharvin resistance~\cite{kumar22}, Poiseuille flow in a channel~\cite{torre15,sulpizio19,ku20,vool21,huang23}, and Stokes flow around obstacles~\cite{lucas17,gusev20}.
One of the most remarkable predictions of hydrodynamic theory is the formation of stationary vortices (or whirlpools)~\cite{levitov16, pellegrino16,falkovich17,guerrerobecerra19, danz20, nazaryan21}, which has been indirectly confirmed by negative resistance measurements caused by current backflow~\cite{bandurin16,braem18,bandurin18}.
Recently, para-hydrodynamic vortices were shown to exist in WTe$_2$ at cryogenic temperatures through direct imaging~\cite{aharonsteinberg22}. Although transport in this system is described by a hydrodynamic theory, the observed vortices do not originate from electron-electron interactions. Genuine electron-hydrodynamic vortices, although widely anticipated~\cite{levitov16,nazaryan21}, have remained challenging to realize.

Here, we demonstrate direct imaging of stationary current whirlpools in a monolayer graphene (MLG) device at room temperature via scanning nitrogen-vacancy (NV) magnetometry (Fig.~\ref{fig1}A).  We study the crossover regime from vortex-free to vortex flow (presence of a single whirlpool). We find that the vortex signature is most pronounced in the smallest devices and disappears upon increasing the device size. We observed the whirlpools in both electron and hole-dominated regimes, but not as the doping approached charge neutrality.  Overall, our measurements are well explained by a hydrodynamic description and clearly rule out a purely diffusive theory.

\subsection*{Imaging of current whirlpools}

The collective motion of a viscous electron fluid can be described by the Navier-Stokes equation in conjunction with the continuity equation~\cite{torre15, lucas18},
\begin{align}
  \vec{J}(\vec{r}) - \Dv^2\nabla^2\vec{J}(\vec{r}) &+ \sigma_0\nabla\phi(\vec{r}) = 0 \ , \label{eq:navierstokes} \\
  \nabla\cdot \vec{J}(\vec{r}) &= 0 \ . \label{eq:continuity}
\end{align}
Here, the current density $\vecJ(\vecr)$ reflects the flow velocity subject to a potential gradient $\nabla\phi(\vecr)$ and a viscous term $\nabla^2\vecJ$.  $\Dv$ is the characteristic length scale describing vorticity diffusion, commonly referred to as the Gurzhi length, and $\sigma_0$ is the Drude conductivity~\cite{torre15}.
The Gurzhi length can further be related to microscopic scattering theory via~\cite{guo17,pellegrino17}:
\begin{align}
\Dv = \frac12 (\lmr\lee)^{1/2} \ .
\label{eq:gurzhi}
\end{align}
To resolve spatial signatures of viscous electron flow, the characteristic size of the device should be of similar size or smaller than the Gurzhi length.  For high-quality MLG at room temperature, $\lee$ is of the order of $0.2\unit{\um}$~\cite{principi16,kim20} and $\lmr\sim1.0\unit{\um}$~\cite{wang13}, resulting in an expected $\Dv$ on the order of $0.2\unit{\um}$. 

\begin{figure}[h!]
\center
\includegraphics[width=0.7\linewidth]{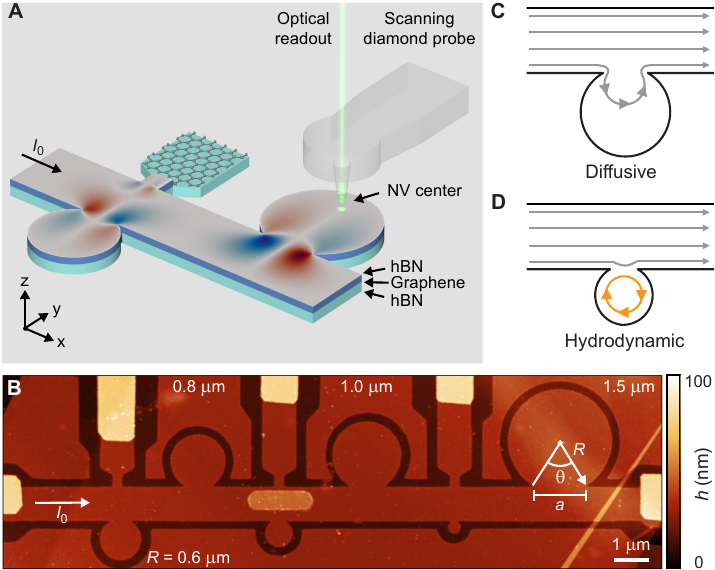}
\caption{{\bf Schematic of the scanning experiment.}
	({\bf A}) Configuration of the encapsulated monolayer graphene (hBN-MLG-hBN) device and scanning nitrogen-vacancy magnetometer. hBN, hexagonal boron nitride.
	({\bf B}) Topography (atomic force microscopy) image of the investigated graphene device.  The device consists of a main channel and disk-shaped side pockets of varying radius $R$.  The disk opening is approximately $a\approx R$ ($\theta\approx 60^\circ$ by design). Bright features are Au contacts. $\Isd$ is the source-drain current.
	({\bf C}) Schematic of current flow in the diffusive regime. 
	({\bf D}) In the hydrodynamic regime, current flow inside the disk reverses direction.}
\label{fig1}
\end{figure}

Our MLG device consists of a uniform channel with disk-shaped side pockets (Fig.~\ref{fig1}B).  For this geometry, the critical length scale is mostly set by the disk opening $a$~\cite{aharonsteinberg22}.  When $a$ is much larger than $\Dv$, the channel current can enter the disk and produce a co-flowing current inside the disk (Fig.~\ref{fig1}C).  The flow pattern is primarily governed by the potential gradient $\nabla\phi(\vecr)$ and resembles diffusive transport.  By contrast, when the disk opening is similar to or smaller than $\Dv$, the laminar current through the main channel can no longer enter the disk; instead, a counter-flowing vortex current appears mediated by momentum-conserving interactions (Fig.~\ref{fig1}D).  Therefore, the current direction in the disk -- co-flowing or counter-flowing -- serves as a hallmark to discriminate between diffusive and hydrodynamic transport.

To map the current distribution in the channel and disk, we image the current-generated magnetic field $\sim 70\unit{nm}$ above the MLG sheet using a scanning NV magnetometer~\cite{chang17} (Fig.~\ref{fig1}A).  We use current amplitudes $I_0 = 2-30\unit{\mu A}$, which are sufficiently small to not heat the electron gas but still easily detectable by our magnetometer~\cite{palm22}.  To further enhance the sensitivity, we modulate the current at $f\sim 25-65\unit{kHz}$ and synchronize it with a spin-echo detection of the spin sensor's quantum phase~\cite{ku20,palm22}.  A graphite back gate located $\sim 24\ \unit{nm}$ beneath the graphene flake is used to tune the carrier type (electrons, holes) and concentration between ca. $\pm 2\ee{12}\unit{cm^{-2}}$.

Even deep into the hydrodynamic regime, the vortex current is expected to reach only a few percent of the total current $I_0$.  To discern the subtle vortex texture from the dominating channel flow, we align the device such that the channel current flows along $x$ while the transverse currents in and out of the disk flow along $y$.  Consequently, we can use the two magnetic field components $\Bx \sim +\mo\JyTilde/2$ and $\By \sim -\mo\JxTilde/2$ to obtain separate maps for each current direction.  Here, $\JxTilde$ and $\JyTilde$ are the low-pass-filtered (due to the NV standoff distance) sheet current densities with units of ampere per meter; $\mo=4\pi\ee{-7}\unit{T/(Am^{-1})}$; see \cite{supplemental} for a discussion of the current reconstruction.

\begin{figure}[h!]
\center
\includegraphics[width=0.7\linewidth]{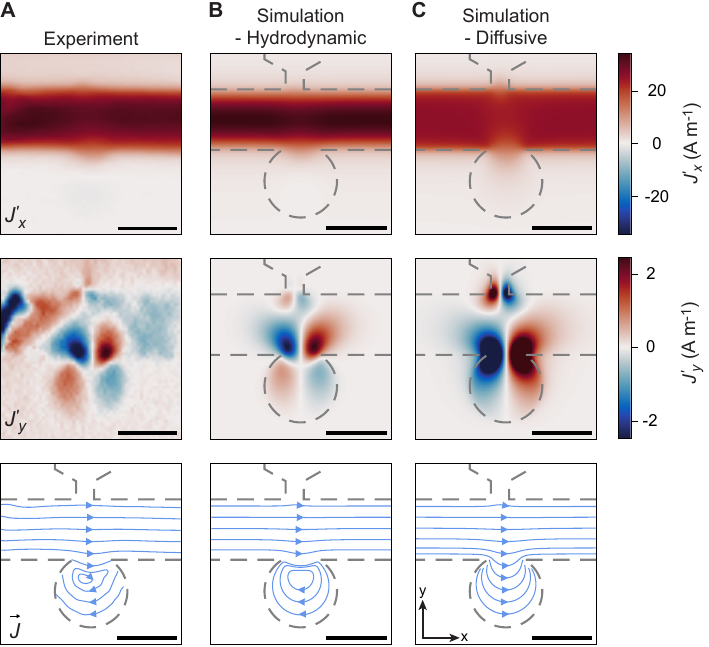}
\caption{{\bf Observation of current whirlpools}.
	({\bf A}) Measured channel flow $\JxTilde$ (top), transverse flow $\JyTilde$ (middle), and velocity plot of the current density vector $\vecJ$ (bottom) in the hole-doped regime ($n\approx-1.7\ee{12}\unit{cm^{-2}}$).
	({\bf B}) Simulation of the same geometry using the hydrodynamic model ($\Dv=0.28\unit{\um}$).
	({\bf C}) Simulation using the diffusive model ($\Dv=0.001\unit{\um}$).
	Both simulations use a no-slip boundary condition. Simulated maps are low-pass filtered for direct comparison with the experimental $\JxTilde$ and $\JyTilde$ maps \cite{supplemental}. The dashed lines indicate the device edges.  Scale bars are $1\unit{\um}$. Measurements were taken at room temperature.}
\label{fig2}
\end{figure}

Figure~\ref{fig2}A shows experimental maps of the current flow in the $R=0.6\unit{\um}$ disk, together with numerical simulations of Eqs.~\ref{eq:navierstokes} and \ref{eq:continuity} for the hydrodynamic case (Fig.~\ref{fig2}B) and the diffusive case (Fig.~\ref{fig2}C), respectively.
The sign and shape of the measured $\JyTilde$ matches the counter-flow of the viscous simulation. In addition to the vortex feature in the $R=0.6\unit{\um}$ disk, the experiment also reproduces the smaller current vortex in the lateral voltage probe and the reduction in $\JxTilde$ along the channel edges [indicative of Poiseuille flow; see \cite{supplemental}]. The hallmark sign of $\JyTilde$ and the detailed agreement between simulated and experimental maps constitute the first piece of evidence that transport is governed by electron hydrodynamics in our doped MLG device.

\subsection*{Transition from viscosity to diffusion-dominated transport}

\begin{figure}[h!]
\center
\includegraphics[width=0.6\linewidth]{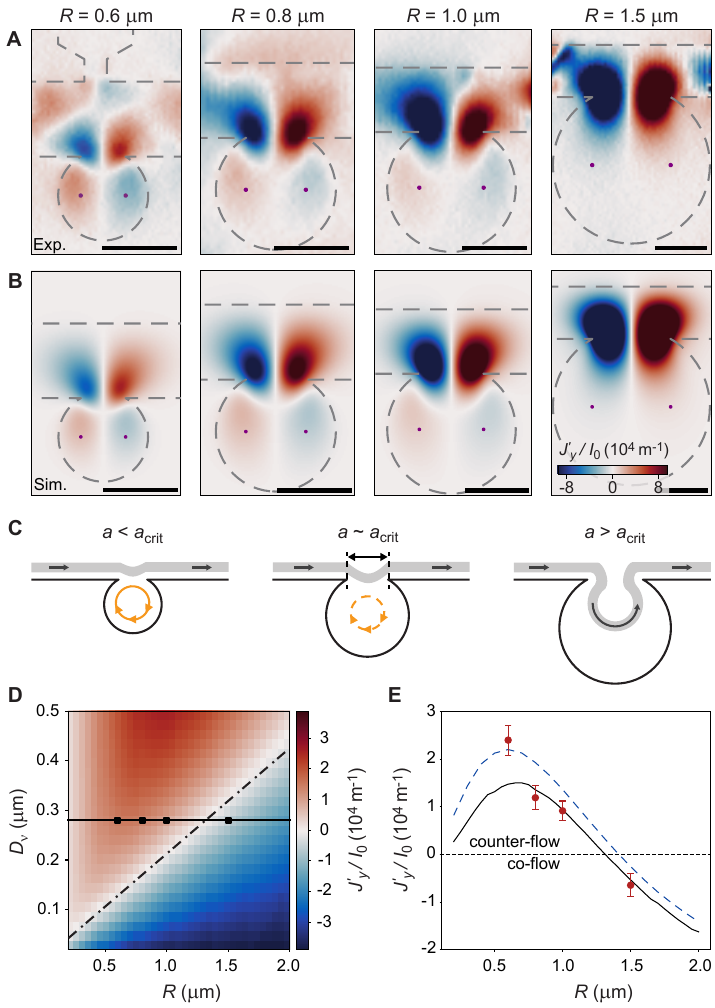}
\caption{{\bf Disk size determines the transport regime.}
	({\bf A} and {\bf B}) Transverse flow $\JyTilde$ as a function of disk radius $R$. Top row (A) shows the experimental data and bottom row (B) shows the simulation using $\Dv=0.28\unit{\um}$ with a no-slip boundary condition. All plots are normalized by the device current $I_0$.
	({\bf C}) Schematic illustrating the transition from vortex flow to vortex-free flow. 
	({\bf D}) Magnitude of the backflow as a function of disk size and Gurzhi length (numerical simulation).Plotted is the transverse current density $\JyTilde:=[\JyTilde(-R/2,0)-\JyTilde(R/2,0)]/2$ at locations $(\pm R/2, 0)$ relative to the center of the disk, marked by dots in (A) and (B).	The black squares are from the simulations in (B). The horizontal center line corresponds to $\Dv = 0.28\unit{\um}$. The dash-dotted line indicates the critical device size $R_{\mr{crit}}\approx a_{\mr{crit}}$ where $\JyTilde$ changes sign.
	({\bf E}) Transverse current density $\JyTilde$ plotted as a function of $R$.  Red dots are the experimental data extracted from the maps in (A) (error bars are two standard deviations). Curves correspond to simulations using $\Dv = 0.28\unit{\um}$ assuming a no-slip boundary condition (solid black line) and a finite slip length ($\lb = 81\unit{nm}$, blue dashed line), respectively \cite{supplemental}. Measurements were taken at room temperature.}
\label{fig3}
\end{figure}

To further support the hydrodynamic model, we image current flow in several disks ($R=0.6-1.5\unit{\um}$) at a fixed carrier density of $n\approx-1.7\ee{12}\unit{cm^{-2}}$, shown in Fig.~\ref{fig3}A.  Vortices are present up to $R=1.0\unit{\um}$ and absent for the largest disk ($R=1.5\unit{\um}$), indicating the transition out of a viscosity-dominated transport regime.  Assuming a device-independent Gurzhi length of $\Dv=0.28\unit{\um}$, we accurately reproduce this transition with numerical simulations (Fig.~\ref{fig3}B).

The disappearance of the vortex with larger disk size may be explained with an intuitive picture (Fig.~\ref{fig3}C): as $R$ increases, so does the disk opening $a \approx R$ (see Fig.~\ref{fig1}B).  When $a$ is small, the channel current cannot enter the disk because viscosity suppresses the in- and out-flowing currents; meanwhile, a vortex is generated in the disk through momentum transfer (left sketch).  As $a$ approaches the critical opening $\acrit\approx 4.7\Dv$~\cite{aharonsteinberg22}, current starts entering the disk and the vortex fades (middle). Above $\acrit$, the disk current reverses direction and flows as is expected from diffusive transport (right).  Because the flow pattern depends on the ratio $a/\Dv$, we can estimate $D_{\nu}$ by plotting the normalized transverse current density extracted symmetrically around the disk center as a function of $R \approx a$ (Fig.~\ref{fig3}, D and E).
Whereas we find excellent agreement for the larger disks, our model underestimates the vortex flow for the smallest disk ($R=0.6\unit{\um}$).  The deviation is likely caused by the assumption of a no-slip boundary condition; refined simulations with a finite slip length and complementary lattice Boltzmann simulations both predict increased counter-flow for smaller disks (Fig.~S13).

\subsection*{Hole and electron carriers}

\begin{figure}[h!]
\center
\includegraphics[width=0.7\linewidth]{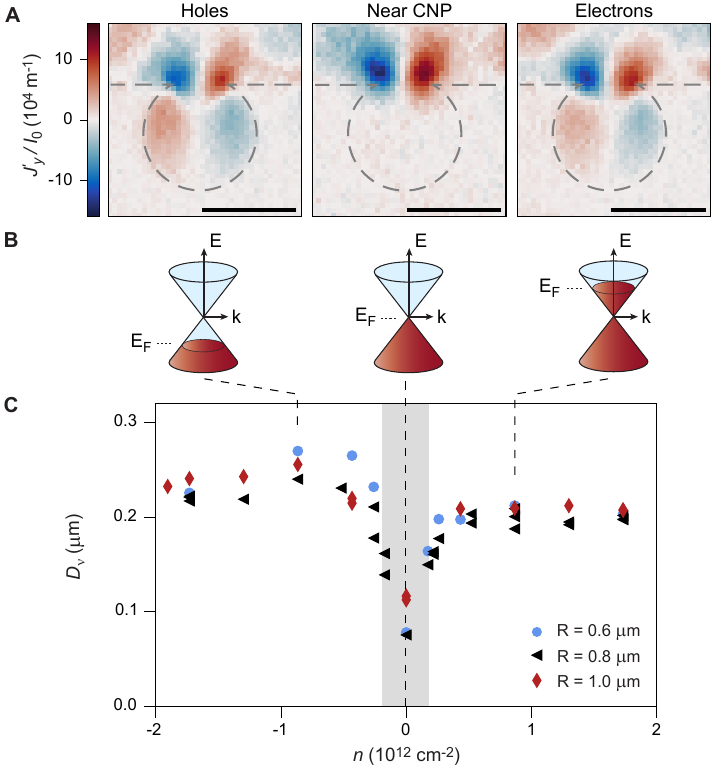}
\caption{{\bf Carrier dependence of the Gurzhi length.}
	({\bf A}) Experimental $\JyTilde$ flow for hole doping at $n\approx -0.9\cdot 10^{12}\unit{cm^{-2}}$\ (left), near charge neutrality (middle), and for electron doping at $n\approx 0.9\cdot 10^{12}\unit{cm^{-2}}$\ (right) for the $R=0.6\unit{\um}$ disk. Scale bars are $1\unit{\um}$.
	({\bf B}) Schematic representation of the electronic band structure and location of the Fermi energy $\EF$ for the scans shown in (A).
	({\bf C}) Gurzhi length $\Dv$ as a function of carrier density $n$. Corresponding plots for the slip length $\lb$ and fits of the channel flow profiles are shown in Figs.~S3 and S4, respectively. The gray region indicates the ambipolar transport regime [$|\EF| \le 2\kB T$, where $\kB$ is the Boltzmann constant; see \cite{supplemental}]. The uncertainties of $\Dv$ values are around $\pm 0.025\unit{\um}$, and dominated by systematic errors caused by an imprecise knowledge of the device geometry [see \cite{supplemental} and Fig.~S6]. Measurements were taken at room temperature.}
\label{fig4}
\end{figure}

We next turn our attention to the carrier density dependence of the vortex flow. Transport models for graphene predict that both $\lmr$ and $\lee$ vary with carrier density \cite{sarma11,li13,ho18}, thus $\Dv \propto \sqrt{\lee \lmr}$ should also depend on $n$.  Figure~\ref{fig4}A shows flow patterns for the $R=0.6\unit{\um}$ disk recorded for hole doping at $n\approx -0.9\ee{12}\unit{cm^{-2}}$, near the charge neutrality point (CNP), and for electron doping at $n\approx0.9\ee{12}\unit{cm^{-2}}$.  Vortex flow is observed in both hole-dominated and electron-dominated regimes.  Notably, however, the current backflow disappears near charge neutrality.

For a more quantitative analysis, we record a series of magnetic field maps for varying carrier densities and fit them with numerical simulations to extract values for $\Dv$.  Details regarding these simulations, including the implementation of a finite slip length boundary condition \cite{torre15,kiselev19}, are discussed in \cite{supplemental}. The resulting values for $\Dv$ are plotted as a function of $n$ in Fig.~\ref{fig4}C.  The data show a strong reduction of $\Dv$ near the CNP; $\Dv$ is approximately constant away from charge neutrality.  Consistent with previous observations \cite{ku20,huang23}, we further observe a slight tendency for $\Dv$ to decrease for large (hole) doping.  Note that around the CNP, the data are still best described by a hydrodynamic model with non-vanishing $\Dv$, as opposed to a fully diffusive model (Fig.~S15).

The strong reduction of the Gurzhi length $\Dv$ near the CNP, which has also been observed in a previous imaging experiment~\cite{jenkins22}, can be explained by a reduction of the microscopic scattering lengths.  In the low-density Fermi liquid regime near the CNP, charged impurity scattering is likely to limit the conductivity in our device ($\sigma_0 \propto n$) \cite{dean10,sarma11}.  Consequently, the mean free path with respect to momentum-relaxing interactions $\lmr = \frac{h}{2e^2}\frac{\sigma_0}{\sqrt{\pi n}}$ becomes proportional to $\sqrt{n}$.  Furthermore, $\lee$ scales approximately as $\sqrt{n}$ \cite{kim20,bandurin18}.  Therefore, $D_{\nu}$ is expected to increase with carrier density near charge neutrality.  In the ambipolar regime, current-relaxing electron-hole collisions need to be accounted for \cite{nam17,fritz23}, and more elaborate transport models may be required to  describe the electronic transport accurately \cite{narozhny21} and to connect the fitted values for $D_{\nu}$ to the microscopic scattering lengths.

Curiously, we find that $\Dv$ is slightly larger for holes compared to electrons.  This carrier asymmetry is also evident by a mildly increased vortex flow for holes in the $R=0.6\unit{\um}$ disk (Fig.~\ref{fig4}A).  In addition, we observe an electron-hole inequality in the smallest investigated structure ($R=0.2\unit{\um}$, Fig.~S14).
Further evidence for a carrier asymmetry is provided by a fit to the current flow profile along the main channel, which is expected to follow the Poiseuille law.  Interestingly, these fits yield $\Dv$ values for holes that are almost one-half the size of the vortex fits (Fig.~S4A).  By contrast, $\Dv$ values for electrons are similar to those extracted from the vortex fits.
Such electron-hole asymmetries are not expected from theory and merit further investigation.
A possible explanation is a carrier-type-dependent doping at the device edge, which would manifest itself in modified boundary conditions \cite{barnard17}.

\subsection*{Discussion and outlook}

Our experiments demonstrate that hydrodynamic whirlpools mediated by electron-electron interactions can be observed in high-mobility materials where $\lmr>\lee$.  The reversal of the current direction provides a clear spatial hallmark of hydrodynamic transport compared to other signatures such as Poiseuille flow~\cite{ku20}.  Additionally, unlike the intermediate temperatures ($T\lesssim 200\unit{K}$) required to observe hydrodynamic flow through a constriction~\cite{jenkins22}, we find clear hydrodynamic signatures at room temperature, likely because of our smaller device geometry.

Although vortex-like features can also emerge in the ballistic regime~\cite{nazaryan21,aharonsteinberg22}, this is unlikely in our case for several reasons: first, to be dominated by ballistic effects, $\lee$ would need to be comparable or larger than the disk diameter, which is $2R\approx 2\unit{\um}$ for the largest disk where we observe a current whirlpool (Fig.~\ref{fig3}A).  This value is an order of magnitude larger than previously reported $\lee\sim 0.1-0.25\unit{\um}$ at room temperature~\cite{kim20,huang23}. Second, vortex flow patterns in the ballistic regime, although possible~\cite{nazaryan21,aharonsteinberg22}, are expected to deviate from those predicted by the hydrodynamic model.  Yet, we observe detailed agreement between our experimental data and the hydrodynamic simulation (Fig.~\ref{fig3}).  Because the transition from the hydrodynamic to the ballistic regime is smooth~\cite{sulpizio19,nazaryan21}, however, a minor ballistic contribution to the flow pattern cannot be ruled out for the smallest disks ($R \lesssim 0.6\unit{\um}$).

Further studies will be needed to investigate the nature of boundary scattering in more detail, especially in view of the observed electron-hole asymmetry.  Our data suggest that some edge defects may only affect transport for a single carrier type (Fig.~S14), potentially because of edge doping~\cite{barnard17}.  More work is required to gauge whether a simple boundary condition using a single parameter (the slip length $\lb$) is sufficient to describe these effects. Corresponding experimental studies would benefit from lower temperatures where the slip length is larger~\cite{kiselev19}, or a smaller device size where boundary effects are more prominent.
Beyond graphene monolayers, bilayer graphene (BLG) is a next obvious candidate, as the steeper rise of $\lee$ with carrier density~\cite{bandurin18,ho18}, lower viscosity~\cite{bandurin16}, and potentially dominant electron-hole collisions near charge neutrality~\cite{nam17} prominently alter the transport physics.  Although BLG has been shown to exhibit a hydrodynamic transport regime~\cite{bandurin16,bandurin18}, it has thus far eluded verification through scanning methods~\cite{palm22}. 
Finally, an exciting prospect is the imaging of non-linear hydrodynamic effects, such as preturbulence~\cite{mendoza11,gabbana18} and turbulence~\cite{disante20}, which may be possible with NV centers via relaxometry measurements~\cite{kolkowitz15,ariyaratne18}.


\bibliographystyle{Science}

\begin{thebibliography}{10}

\bibitem{chen16science}
S.~W. Chen, {\it et~al.\/}, {\it Science\/} {\bf 353}, 1522 (2016).

\bibitem{vanwees88}
B.~J. van Wees, {\it et~al.\/}, {\it Physical Review Letters\/} {\bf 60}, 848
  (1988).

\bibitem{lucas18}
A.~Lucas, K.~C. Fong, {\it Journal of Physics: Condensed Matter\/} {\bf 30},
  053001 (2018).

\bibitem{narozhny22}
B.~N. Narozhny, {\it La Rivista del Nuovo Cimento\/} {\bf 45}, 661 (2022).

\bibitem{levitov16}
L.~Levitov, G.~Falkovich, {\it Nat. Phys.\/} {\bf 12}, 672 (2016).

\bibitem{disante20}
D.~D. Sante, {\it et~al.\/}, {\it Nature Communications\/} {\bf 11}, 3997
  (2020).

\bibitem{gurzhi68}
R.~N. Gurzhi, {\it Soviet Physics Uspekhi\/} {\bf 11}, 255 (1968).

\bibitem{molenkamp94}
L.~W. Molenkamp, M.~J.~M. de~Jong, {\it Physical Review B\/} {\bf 49}, 5038
  (1994).

\bibitem{dejong95}
M.~J.~M. de~Jong, L.~W. Molenkamp, {\it Phys. Rev. B\/} {\bf 51}, 13389 (1995).

\bibitem{berdyugin19}
A.~I. Berdyugin, {\it et~al.\/}, {\it Science\/} {\bf 364}, 162 (2019).

\bibitem{kim20}
M.~Kim, {\it et~al.\/}, {\it Nature Communications\/} {\bf 11}, 2339 (2020).

\bibitem{scaffidi17}
T.~Scaffidi, N.~Nandi, B.~Schmidt, A.~P. Mackenzie, J.~E. Moore, {\it Physical
  Review Letters\/} {\bf 118}, 226601 (2017).

\bibitem{pellegrino17}
F.~M.~D. Pellegrino, I.~Torre, M.~Polini, {\it Physical Review B\/} {\bf 96},
  195401 (2017).

\bibitem{guo17}
H.~Guo, E.~Ilseven, G.~Falkovich, L.~S. Levitov, {\it Proceedings of the
  National Academy of Sciences\/} {\bf 114}, 3068 (2017).

\bibitem{krishnakumar17}
R.~Krishna~Kumar, {\it et~al.\/}, {\it Nature Physics\/} {\bf 13}, 1182 (2017).

\bibitem{ginzburg21}
L.~V. Ginzburg, {\it et~al.\/}, {\it Physical Review Research\/} {\bf 3},
  023033 (2021).

\bibitem{kumar22}
C.~Kumar, {\it et~al.\/}, {\it Nature\/} {\bf 609}, 276 (2022).

\bibitem{torre15}
I.~Torre, A.~Tomadin, A.~K. Geim, M.~Polini, {\it Phys. Rev. B\/} {\bf 92},
  165433 (2015).

\bibitem{sulpizio19}
J.~A. Sulpizio, {\it et~al.\/}, {\it Nature\/} {\bf 576}, 75 (2019).

\bibitem{ku20}
M.~J.~H. Ku, {\it et~al.\/}, {\it Nature\/} {\bf 583}, 537 (2020).

\bibitem{vool21}
U.~Vool, {\it et~al.\/}, {\it Nature Physics\/} {\bf 17}, 1216 (2021).

\bibitem{huang23}
W.~Huang, {\it et~al.\/}, {\it Phys. Rev. Res.\/} {\bf 5}, 023075 (2023).

\bibitem{lucas17}
A.~Lucas, {\it Phys. Rev. B\/} {\bf 95}, 115425 (2017).

\bibitem{gusev20}
G.~M. Gusev, A.~S. Jaroshevich, A.~D. Levin, Z.~D. Kvon, A.~K. Bakarov, {\it
  Scientific Reports\/} {\bf 10}, 7860 (2020).

\bibitem{pellegrino16}
F.~M.~D. Pellegrino, I.~Torre, A.~K. Geim, M.~Polini, {\it Physical Review B\/}
  {\bf 94}, 155414 (2016).

\bibitem{falkovich17}
G.~Falkovich, L.~Levitov, {\it Physical Review Letters\/} {\bf 119}, 066601
  (2017).

\bibitem{guerrerobecerra19}
K.~A. Guerrero-Becerra, F.~M.~D. Pellegrino, M.~Polini, {\it Physical Review
  B\/} {\bf 99}, 041407 (2019).

\bibitem{danz20}
S.~Danz, B.~N. Narozhny, {\it 2D Materials\/} {\bf 7}, 035001 (2020).

\bibitem{nazaryan21}
K.~G. Nazaryan, L.~Levitov, {\it arXiv:2111.09878\/}  (2021).

\bibitem{bandurin16}
D.~A. Bandurin, {\it et~al.\/}, {\it Science\/} {\bf 351}, 1055 (2016).

\bibitem{braem18}
B.~A. Braem, {\it et~al.\/}, {\it Physical Review B\/} {\bf 98}, 241304 (2018).

\bibitem{bandurin18}
D.~A. Bandurin, {\it et~al.\/}, {\it Nature Communications\/} {\bf 9}, 4533
  (2018).

\bibitem{aharonsteinberg22}
A.~Aharon-Steinberg, {\it et~al.\/}, {\it Nature\/} {\bf 607}, 74 (2022).

\bibitem{principi16}
A.~Principi, G.~Vignale, M.~Carrega, M.~Polini, {\it Physical Review B\/} {\bf
  93}, 125410 (2016).

\bibitem{wang13}
L.~Wang, {\it et~al.\/}, {\it Science\/} {\bf 342}, 614 (2013).

\bibitem{chang17}
K.~Chang, A.~Eichler, J.~Rhensius, L.~Lorenzelli, C.~L. Degen, {\it Nano
  Letters\/} {\bf 17}, 2367 (2017).

\bibitem{palm22}
M.~L. Palm, {\it et~al.\/}, {\it Physical Review Applied\/} {\bf 17}, 054008
  (2022).

\bibitem{supplemental}
{\it See Supplemental Material accompanying this manuscript\/} .

\bibitem{sarma11}
S.~D. Sarma, S.~Adam, E.~H. Hwang, E.~Rossi, {\it Reviews of Modern Physics\/}
  {\bf 83}, 407 (2011).

\bibitem{li13}
Q.~Li, S.~D. Sarma, {\it Physical Review B\/} {\bf 87}, 085406 (2013).

\bibitem{ho18}
D.~Y.~H. Ho, I.~Yudhistira, N.~Chakraborty, S.~Adam, {\it Phys. Rev. B\/} {\bf
  97}, 121404 (2018).

\bibitem{kiselev19}
E.~I. Kiselev, J.~Schmalian, {\it Physical Review B\/} {\bf 99}, 035430 (2019).

\bibitem{jenkins22}
A.~Jenkins, {\it et~al.\/}, {\it Physical Review Letters\/} {\bf 129}, 087701
  (2022).

\bibitem{dean10}
C.~R. Dean, {\it et~al.\/}, {\it Nat. Nanotech.\/} {\bf 5}, 722 (2010).

\bibitem{nam17}
Y.~Nam, D.~Ki, D.~Soler-Delgado, A.~F. Morpurgo, {\it Nature Physics\/} {\bf
  13}, 1207 (2017).

\bibitem{fritz23}
L.~Fritz, T.~Scaffidi, {\it arXiv:2303.14205\/}  (2023).

\bibitem{narozhny21}
B.~N. Narozhny, I.~V. Gornyi, M.~Titov, {\it Physical Review B\/} {\bf 104},
  075443 (2021).

\bibitem{barnard17}
A.~W. Barnard, {\it et~al.\/}, {\it Nature Communications\/} {\bf 8}, 15418
  (2017).

\bibitem{mendoza11}
M.~Mendoza, H.~J. Herrmann, S.~Succi, {\it Physical Review Letters\/} {\bf
  106}, 156601 (2011).

\bibitem{gabbana18}
A.~Gabbana, M.~Polini, S.~Succi, R.~Tripiccione, F.~M.~D. Pellegrino, {\it
  Phys. Rev. Lett.\/} {\bf 121}, 236602 (2018).

\bibitem{kolkowitz15}
S.~Kolkowitz, {\it et~al.\/}, {\it Science\/} {\bf 347}, 1129 (2015).

\bibitem{ariyaratne18}
A.~Ariyaratne, D.~Bluvstein, B.~A. Myers, A.~C.~B. Jayich, {\it Nature
  Communications\/} {\bf 9}, 2406 (2018).

\bibitem{palm24zenodo}
M.~L. Palm, {\it et~al.\/}, {\it Zenodo (2024);
  http://doi.org/10.5281/zenodo.10124549\/} .

\end{thebibliography}

\clearpage
\section*{Acknowledgments}

The authors thank Matthew Markham (ElementSix) for providing $^{12}$C diamond material, Jan Rhensius (QZabre) for nanofabrication, the Ensslin group, the FIRST Lab, and the Euler computer cluster at ETH Zurich for access to their instrumentation, Eli Zeldov, Lev Ginzburg, and Klaus Ensslin for helpful discussions, Alex Eichler and Jan Rhensius for help with the illustration in Figure 1, and Marius Bild for help with the Zenodo repository.
\textbf{Funding:} This work was supported by the European Research Council through ERC CoG 817720 (IMAGINE), the Swiss National Science Foundation (SNSF) through the National Centre of Competence in Research in Quantum Science and Technology (NCCR QSIT), Grant No. 51NF40-185902, and the Advancing Science and TEchnology thRough dIamond Quantum Sensing (ASTERIQS) program, Grant No. 820394, of the European Commission.
K.W. and T.T. acknowledge support from the JSPS KAKENHI (Grant Numbers 20H00354, 21H05233 and 23H02052) and World Premier International Research Center Initiative (WPI), MEXT, Japan. 
\textbf{Author contributions:} C.L.D. and M.L.P. conceived of the experiment. M.L.P., C.D., and W.S.H. carried out experiments. M.L.P. and C.D. performed the data analysis. C.D. and M.L.P. performed macroscopic simulations, and W.S.H. performed  Boltzmann simulations. M.L.P. fabricated the sample. T.T. and K.W. provided the hexagonal boron nitride. M.L.P., C.D., W.S.H., and C.L.D. wrote the manuscript. All authors discussed the results.
\textbf{Competing interests:} The authors have no conflicts to disclose.
\textbf{Data and materials availability:} All data and software are available in the manuscript or the supplementary material or are deposited at Zenodo~\cite{palm24zenodo}.

\section*{Supplementary Materials}
Materials and Methods\\
Supplementary Text 1-5\\
Figs. S1 to S15\\

\clearpage
\section*{Figures and Captions}

\subsection*{Figure 1} 

\clearpage
\subsection*{Figure 2} 

\clearpage
\subsection*{Figure 3}

\clearpage
\subsection*{Figure 4} 

\end{document}


\begin{center}
{\LARGE{}Supplementary Materials for} \\
\vspace{0.5cm}
{\Large{} Observation of current whirlpools in graphene at room temperature} \\
\vspace{0.5cm}
Marius~L.~Palm$^{1\dagger}$, Chaoxin~Ding$^{1\dagger}$, William~S.~Huxter$^{1\dagger}$, \\	Takashi Taniguchi$^{2}$, Kenji Watanabe$^{3}$, and Christian~L.~Degen$^{1,4\ast}$ \\

\vspace{0.5cm}
\normalsize{$^1$Department of Physics, ETH Zurich, Otto Stern Weg 1, 8093 Zurich, Switzerland;} \\
\normalsize{$^2$Research Center for Materials Nanoarchitectonics, National Institute for Materials Science,} \\
\normalsize{1-1 Namiki, Tsukuba 305-0044, Japan;} \\
\normalsize{$^3$Research Center for Electronic and Optical Materials, National Institute for Materials Science,} \\
\normalsize{1-1 Namiki, Tsukuba 305-0044, Japan;} \\
\normalsize{$^4$Quantum Center, ETH Zurich, 8093 Zurich, Switzerland.} \\

\vspace{0.5cm}
$^\ast$ correspondence to: degenc@ethz.ch \\
$^\dagger$ these authors contributed equally to this work. \\
\end{center}

\vspace{1.0cm}
{\bf This PDF file includes:}
\vspace{0.5cm} \\
Materials and Methods \\
Supplementary Text 1 to 5 \\
Figures S1 to S15 \\

\clearpage

\section*{Materials and Methods}

\subsection*{Device fabrication}

\begin{figure}[h!]
\centering
\includegraphics[width=0.5\textwidth]{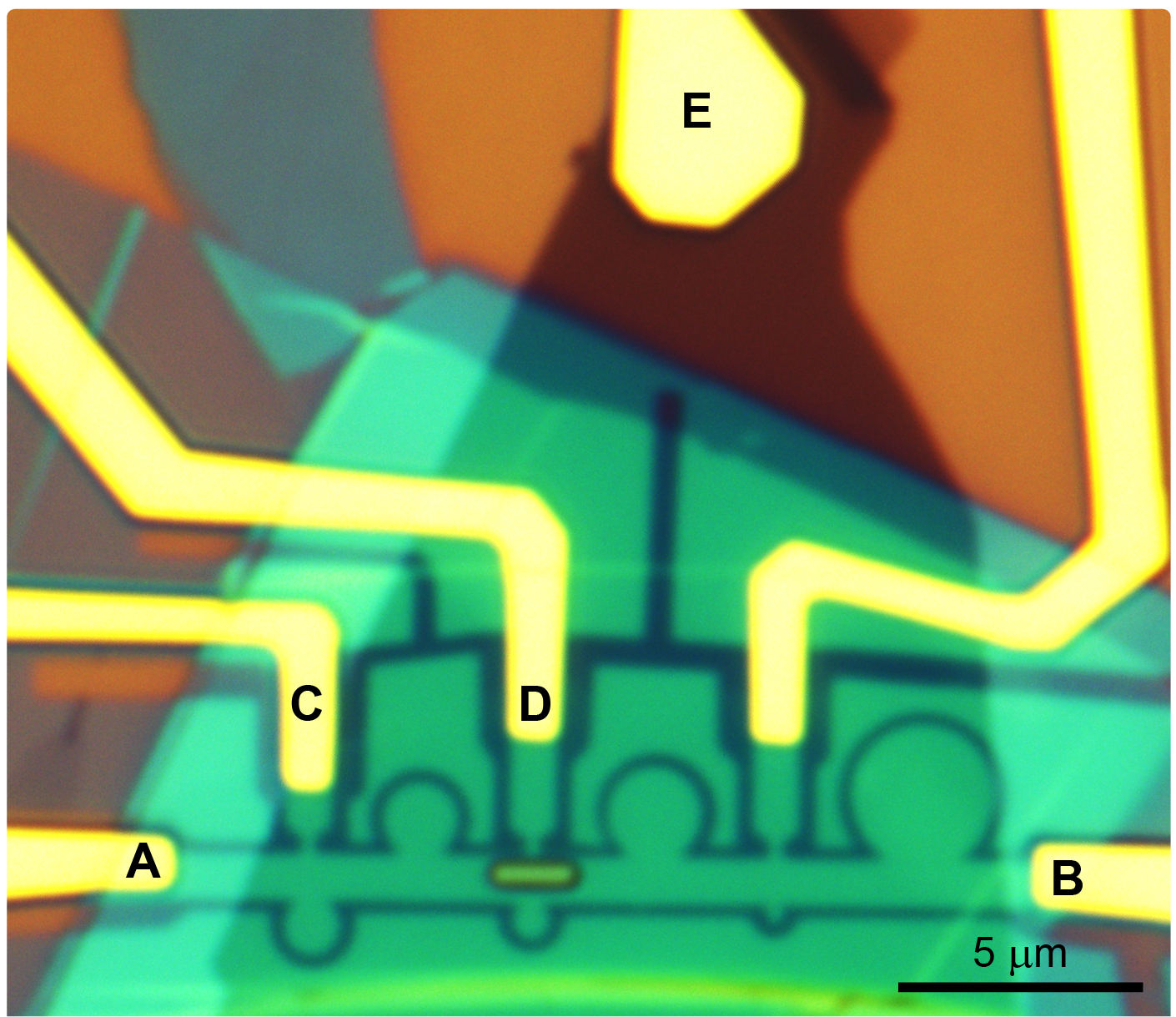}
\caption{\captionstyle{\bf Optical microscope image of the whirlpool device.}
	We send a current through contacts A and B, and use contacts C and D for monitoring the longitudinal voltage drop. The carrier density in the graphene sheet can be tuned via contact E.  The unlabeled contact is not connected.}
\label{fig:suppl:fab}
\end{figure}

The encapsulated graphene device is assembled from mechanically-exfoliated graphene and hexagonal boron nitride flakes using a pick-up technique \cite{wang13, zomer14}. The sample is then annealed in an Ar atmosphere at $350\unit{^{\circ}C}$ for 3 hours. We define the shape of the contacts via electron beam lithography using a triple-layer resist film of AR-P 632.04, AR-P 672.045, and AR-PC 5090.02, etch away the top hBN layer (CHF$_3$/O$_2$ plasma) and create a one-dimensional contact to the graphene sheet through deposition of a Cr/Au (10/50 nm) film and subsequent lift-off~\cite{wang13}. Through a secondary electron beam lithography step followed by dry etching, the device geometry is defined (visible through the color change in the stack in Fig.~\ref{fig:suppl:fab}). A gold patch was added in a third step to fix a fissure of the graphene sheet next to contact D.

The carrier density in the graphene sheet is tuned via a graphite back gate located $d_{\mathrm{BG}}\approx24\unit{nm}$ below the graphene sheet. Assuming a capacitive model with $\epsilon_r \approx 3.76$ \cite{laturia18}, we use $n = \epsilon_0 \epsilon_r \Vbg / (\mathrm{e} d_{\mathrm{BG}}) \approx 8.7\ee{12}\unit{V^{-1}cm^{-2}} \cdot \Vbg$.  The associated Fermi energy is $|\EF| = \hbar\vF\sqrt{\pi|n|}$, where $\vF$ is the Fermi velocity and where the sign of $\EF$ equals the sign of $n$.  Charge neutrality is typically found near $\Vbg = 0\unit{V}$.

\subsection*{Scanning magnetometer setup}

We use commercially available all-diamond scanning probes attached to quartz tuning forks for tip-sample distance control (QZabre). A lock-in amplifier (Zurich Instruments HF2LI) is used to monitor the tuning fork oscillation amplitude and update the target $z$-position of the sample stage (PI P-527.3CL) using a PID controller. Optical initialization and readout of the NV center is achieved with a confocal microscope ($50\unit{\um}$ pinhole) featuring an objective with a numerical aperture of 0.75 (Mitutoyo M Plan Apo HR 50x). We use a custom-built $520\unit{\mr{nm}}$ pulsed diode laser for optical excitation of the NV center. The photoluminescence of the NV center is recorded with a single-photon avalanche photodiode (Excelitas SPCM-AQRH). For the manipulation of the NV spin state, microwave pulses are generated using an IQ mixer (Marki MMIQ-0205HSM) where the local oscillator is provided by a microwave synthesizer (NI Quicksyn FSW-0020). The I and Q signals are generated by an arbitrary waveform generator (Spectrum DN2.663-04). The microwave delivery is accomplished using an Al bond wire positioned several tens of micrometers away from NV. The degeneracy of the $m_S = \pm 1$ sublevels of the ground state of the NV center is lifted with a permanent magnet located beneath the sample stage.  

During magnetometry operation, we use two analog channels of the arbitrary waveform generator to apply the source-drain voltage $\Vsd$ and the back-gate voltage $\Vbg$ synchronously with the pulsed experiments. The resulting device current $I_0$ is amplified using a transimpedance amplifier (FEMTO DHPCA-100) and monitored with the data acquisition module of a digital lock-in amplifier (Zurich instruments MFLI).

\subsection*{Quantum sensing protocol}

We use an AC quantum sensing technique to measure the magnetic field above the sample of interest \cite{ku20, palm22, kotler11}. After initializing the spin state into the $|0\rangle$ state, a $\pi/2$ pulse is applied to create the superposition state $1/\sqrt{2}(|0\rangle + |{-}1\rangle)$. For a duration $\tau/2$, the spin evolves freely and interacts with the applied magnetic field signal $\vec{B}(t)$. For sufficiently small off-axis fields, the NV is only affected by the component $\BNV(t)$ parallel to the symmetry axis of the NV center \cite{rondin14}. After this evolution time, the spin state can be written as $|\psi\rangle = 1/\sqrt{2}(|0\rangle + e^{i\phi(\tau/2)}|{-}1\rangle)$ with $\phi(\tau/2) = \yNV\int_0^{\tau/2}\BNV(t) \mr{d}t$ \cite{degen17}. Here, $\yNV/(2\pi) = 28.02\unit{GHz/T}$ is the gyromagnetic ratio of the NV electronic spin. A subsequent $\pi$ pulse effectively reverses the coupling between the spin and the magnetic field (spin echo), and therefore, the total acquired phase after another evolution time of $\tau /2$ is given by $\phi(\tau) = \yNV\int_0^{\tau/2}\BNV(t) \mr{d}t - \yNV\int_{\tau/2}^{\tau}\BNV(t) \mr{d}t$. For the sinusoidal signals with period $T = \tau$ used throughout this work, this expression evaluates to $\phi(\tau) = \frac{2}{\pi}\yNV \BNV \tau$. A final $\pi/2$ pulse with a phase $\Phi$ relative to the initial microwave pulse converts $\phi$ into a population difference, and a subsequent optical readout yields a PL signal of the form \cite{knowles16, palm22}:
\begin{equation}
	C_\Phi = C_{\mathrm{ref}}^0\left(1-\frac{\epsilon}{2} + \frac{\epsilon e^{-(\tau/T_2)^\alpha}}{2}\cos\left( \gamma_{\mathrm{NV}} \frac{\pi}{2}\tau B_{\mathrm{||}} + \Phi\right)\right)
	\label{eq:theory:phase}
\end{equation}
Here, $C_{\mathrm{ref}}^0$ is the PL signal of the $m_S = 0$ state, $\epsilon$ is the contrast of the NV center, $T_2$ is the dephasing time, and $\alpha$ is a free exponent. The phase $\phi$ is extracted from a set of four measurements ($\Phi \in \{0, \pi/2, \pi, 3\pi/2\}$),
\begin{equation}
	\phi = \mr{arctan2}(C_{3\pi/2} - C_{\pi/2}, C_{0} - C_{\pi}),
\end{equation}
with $\mr{arctan2}$ being the two-argument arctangent function, see Refs.~\cite{ku20, palm22}.

\subsection*{Reconstruction of current density}

Reconstruction of the current density is performed in two steps.  In a first step, we compute the in-plane ($\Bx$ and $\By$) components of the magnetic field from the measured projection $\BNV$.  We carry out the computation in $k$-space~\cite{roth89, chang17, jenkins22},
\begin{align}
\Bxft &= \frac{i\kx\BNVft}{i\ex\kx + i\ey\ky - \ez k} \\
\Byft &= \frac{i\ky\BNVft}{i\ex\kx + i\ey\ky - \ez k} 
\end{align}
where hat symbols denote two-dimensional Fourier transforms, $\veck = (\kx,\ky)$ is the in-plane $k$-space vector and $k=\sqrt{\kx^2+\ky^2}$.  Further, $\vece=(\ex,\ey,\ez)=(\sin\theta\cos\varphi,\sin\theta\sin\varphi,\cos\theta)$ is the unit vector describing the projection axis and $(\theta,\varphi)$ is the known anisotropy axis of the NV center.

\begin{figure*}[h!]
\centering
\includegraphics[width=0.8\textwidth]{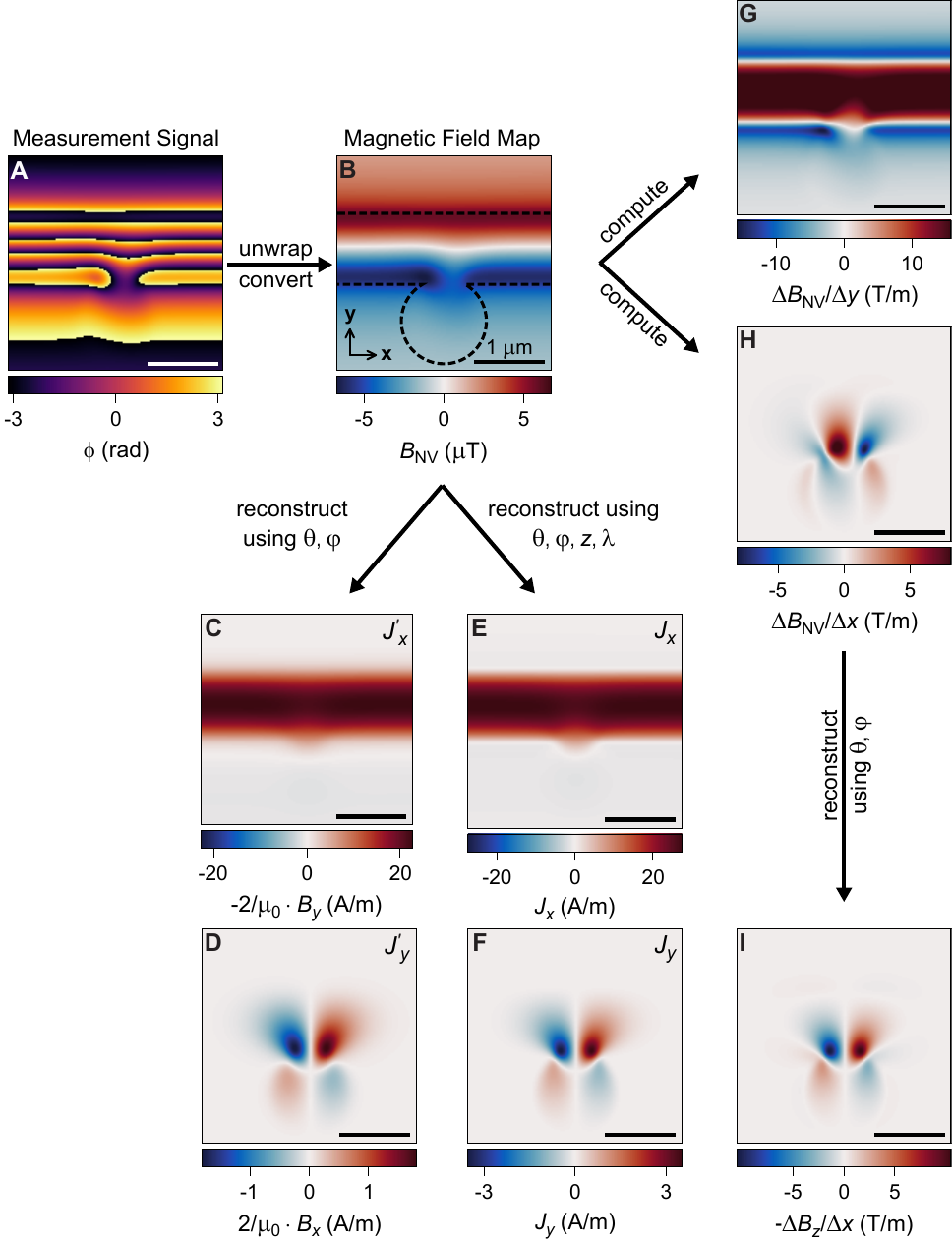}
\caption{\captionstyle{\bf Illustration of the different reconstruction methods and analysis tools used for the investigation of current whirlpools.}
	The data shown in this figure are simulated. Starting from the quantum phase ({\bf A}) as obtained from the Hahn-echo protocol, we first extract the encoded magnetic field map $\BNV$ ({\bf B}) by unwrapping the phase map and using the relation $\phi = \frac{2}{\pi}\gamma_e\BNV \tau$.
	We can then reconstruct the current density $\JxTilde=-\frac{2}{\mu_0}B_y$ and $\JyTilde=\frac{2}{\mu_0}B_x$ ({\bf C} and {\bf D}) using the known NV angles $\theta = 55.7^\circ$, $\varphi = 1^\circ$.
	Alternatively, we can apply inverse filtering \cite{roth89, chang17} to trade signal-to-noise ratio for a slightly improved spatial resolution ($\Jx$ and $\Jy$ in {\bf E} and {\bf F}) using a Hann filter (here with $\lambda = 100\unit{nm}$); however, this inverse filtering was not necessary for most of the data shown in this work. 
	Finally, the signatures from the current flow of the disc can also be disentangled from the channel flow by computing the magnetic field derivatives $\frac{\Delta\BNV}{\Delta y}$ ({\bf G}) and $\frac{\Delta\BNV}{\Delta x}$ ({\bf H}). Via a subsequent computation of $-\frac{\Delta B_z}{\Delta x}$ ({\bf I}) involving the NV angles, a map reminiscent of $J_y$ can be obtained.}
\label{fig:suppl:methodsReconstruction}
\end{figure*}

For recovering the current density vector $\vecJ = (\Jx,\Jy)$, we note that the stray fields in $k$-space are given by
\begin{align}
\Bxft &= \frac12\mo e^{-kz}\Jyft  \\
\Byft &= -\frac12\mo e^{-kz}\Jxft
\end{align}
where $z$ is the standoff distance.  Thus, $\Bx$ and $\By$ are low-pass filtered images of $\Jy$ and $-\Jx$, respectively (see Fig.~\ref{fig:suppl:methodsReconstruction}).  The filter convolution function is given by the inverse Fourier transform of $e^{-kz}$, which has a Lorentzian-like shape,
\begin{align}
G = \mathcal{F}^{-1}\left[e^{-kz}\right] = \frac{z}{2\pi[x^2+y^2+z^2]^{3/2}}
\end{align}
The $\JxTilde$ and $\JyTilde$ maps shown in the main manuscript represent these low-pass filtered maps of $\Jx$ and $\Jy$,
\begin{align}
\JxTilde &= G\ast\Jx = -2\By/\mo \\
\JyTilde &= G\ast\Jy =  2\Bx/\mo
\end{align}
The convolution function has a full width at half maximum of approximately $1.5z$, and sets the minimum feature size for the $\JxTilde$, $\JyTilde$ maps.
Since the spatial transport features in our experiments are typically larger than $100\unit{nm}$, the low-pass filtering only has a minor effect on the images.  If desired, the spatial resolution could be improved to $0.5-1.0 z$ using inverse filtering~\cite{roth89,chang17}, however, this procedure can introduce image artifacts and thus, we refrained from using it in our analysis unless noted otherwise.

Note that due to a singularity at $k = 0$, the offsets of $\JxTilde$ and $\JyTilde$ are undefined \cite{roth89, broadway20}. For the $\JyTilde$-maps presented in this work, we fix the offset by subtracting the average value of the image. Since the channel of the device is oriented along the $x$-axis, we expect approximately equal positive and negative contributions to the $\JyTilde$ image (see Fig.~\ref{fig:suppl:methodsReconstruction}). The aforementioned offset calibration is therefore appropriate for this component of the current density. For the $\JxTilde$ image, an analogous offset correction is not possible and we determine the offset from a region far away from the device.

\subsection*{Simulation of current density maps}

For the simulations presented in the main text, we generally assume that variations in the carrier density can be neglected, and that only a single carrier type is present in the device. Furthermore, we neglect the effects of the small magnetic field (few tens of mT) applied to split the $m_S=\pm1$ sublevels of the NV center, see Supplementary Text~4 for a detailed discussion about the effect of the bias field on the transport. For direct comparison between experimental and simulated data, we compute the low-pass filtered version of the current density ($\JxTilde$, $\JyTilde$) where necessary. 

\subsubsection*{No-slip boundary condition}

For the simulations involving a no-slip boundary condition~\cite{torre15}, we solve the partial differential equation describing the electronic transport using the Partial Differential Equation Toolbox\textsuperscript{TM} in MATLAB\textsuperscript{\textregistered}. We solve Eqs.~(1,2) of the main text for $\left(\phi, \Jx, \Jy \right)$, where $\phi$ is the electric potential, by applying suitable Dirichlet boundary conditions for all boundaries. We note that fixing the electric potential on both the source and drain contact results in varying amounts of current flow depending on simulation parameters such as $\Dv$ and $\sigma_0$. Where necessary, we rescale the results to reflect the desired amount of current flowing through the device. This is possible due to the linearity of the equation and is equivalent to a change of the source-drain potential. All simulations are performed with a mesh size smaller or equal to $20\unit{nm}$.  For the diffusive case, we set $\Dv = 1\unit{nm}$.

\subsubsection*{General boundary condition}

For the experimental determination of the hydrodynamic model parameters, we solve the Navier-Stokes equation using COMSOL Multiphysics\textsuperscript{\textregistered}, similar to Ref.~\cite{aharonsteinberg22}. We solve the partial differential equations for the variables $\left(\phi, \Jx, \Jy\right)$ with a maximal mesh size of $20\unit{nm}$. For the source and drain contacts, we impose Dirichlet boundary conditions fixing the injected current and the potential, respectively. For the remaining boundaries, we impose a Neumann boundary condition for the current density:
\begin{equation}
	-(\vec{n}\cdot\nabla) \vec{J} = -(\vec{n}\cdot\nabla) \vec{J}^t - (\vec{n}\cdot\nabla) \vec{J}^n = \left(l_b^t\right)^{-1}\vec{J}^{t} + \left(l_b^n\right)^{-1}\vec{J}^{n}
\end{equation}
Here, the tangential and normal components of the current density are denoted by the superscripts $t$ and $n$, respectively, where $\vec{n}$ denotes the outward normal vector. We introduce two independent slip length parameters $l_b^t$ and $l_b^n$ for the tangential and the normal component, respectively. By setting $l_b^n = 0.1\unit{pm}$, we force the normal current density $\vec{J}^n$ to (virtually) vanish at the device edge without forcing its derivative $(\vec{n}\cdot\nabla) \vec{J}^n$ to vanish. The remaining tangential part of the boundary condition is identical to the commonly employed boundary condition with variable slip length $l_b^t$ \cite{torre15, kiselev19}:
\begin{equation}
	-(\vec{n}\cdot\nabla) \vec{J}^t = \left(l_b^t\right)^{-1}\vec{J}^{t}
\end{equation}

\subsection*{Estimation of $\Dv$, $\lb$ and $z$ from the vortex flow}

We extract estimates for the vorticity diffusion length $\Dv$, slip length $\lb := l_b^t$ and standoff $z$ by comparing the experimental data to simulations based on the Navier-Stokes equation. For a discrete set of parameters ($D_{\nu} \in \left[10\unit{nm}; 400\unit{nm}\right]$ in steps of $10\unit{nm}$ (up to $500\unit{nm}$ for the $R=0.6\unit{um}$ disc); $l_b \in \left[1\unit{nm}; 193\unit{nm}\right]$ in steps of $8\unit{nm}$), we simulate maps of the current density using COMSOL Multiphysics\textsuperscript{\textregistered} and compute the corresponding magnetic field maps according to Ref.~\cite{roth89}, accounting for a $1\unit{^{\circ}}$ rotation of the sample with respect to the scan axes. We generate magnetic field maps for standoff distances between $50\unit{nm}$ and $120\unit{nm}$ in steps of $2\unit{nm}$. For the computation of the magnetic field projection $\BNV$, we use the NV angles ($\theta \approx 55\unit{^{\circ}}$, $\varphi \approx 1\unit{^{\circ}}$).

\begin{figure}[h!]
\includegraphics[width=0.9\textwidth]{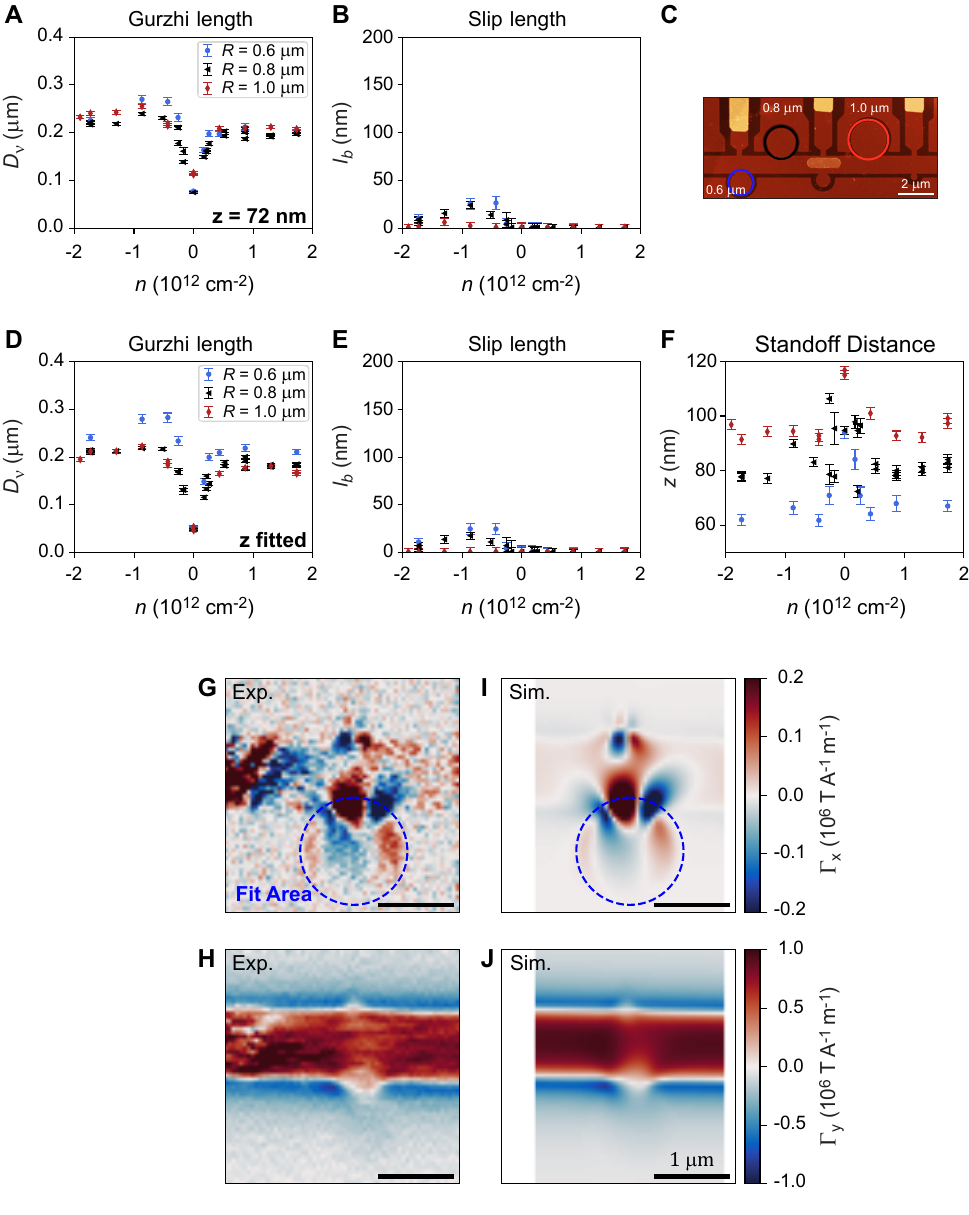}
\caption{\captionstyle{\bf Parameter fits to vortex flow.}
	({\bf A} and \textbf{B}) Carrier density dependence of $D_{\nu}$ and $l_b$ obtained by fitting $\Gamma_x = \frac{1}{I_0} \frac{\Delta B_{\mathrm{NV}}}{\Delta x}$ for several scans on three separate discs. The standoff distance is fixed at $z = 72\unit{nm}$. The fit areas are indicated in {\bf C}. Error bars represent one standard deviation.
	({\bf D-F}) Fit results obtained by optimizing also with respect to the standoff distance $z$. 
	({\bf G} and {\bf H}) Experimental data ($\Gamma_x$, $\Gamma_y$ computed from $\BNV$) taken at $n \approx 0.9\ee{12}\unit{cm^{-2}}$ ($\Vbg = 1\unit{V}$) and ({\bf I} and {\bf J}) simulated maps with $(D_{\nu}, l_b, z=72\unit{nm})$ chosen as close as possible to the fitted parameters. The fit area is indicated by the blue dashed line.
	}
\label{fig:suppl:fittingMethodsDisc}
\end{figure}

For the data shown in Fig.~4C of the main text, we estimate the Gurzhi length and the slip length of the experimental data by fitting the maps of the normalized magnetic field derivative $\Gamma_x = \frac{1}{I_0}\frac{\Delta \BNV}{\Delta x}$ to the hydrodynamic model via nonlinear least squares.
We compare $\Gamma_x$ rather than $B_{\mathrm{NV}}$, since the spatial derivative allows us to disentangle the disc flow more easily from the channel flow (see Fig.~\ref{fig:suppl:methodsReconstruction}). Furthermore, the derivative conveniently removes long-range magnetic field signals originating from current flow in nearby metallic leads. Note that we compare solely the pixels in a circular area with radius $R + 0.1\unit{\um}$ around the disc center (see Fig.~\ref{fig:suppl:fittingMethodsDisc}~(C)). This ensures that the parameter estimation is based on the signatures from the whirlpool and is not affected by imperfections in the channel that are not accounted for by the model. We use cubic interpolation to generate maps of $\Gamma_x$ for fit parameters not covered by our discrete set of simulations. The results of this fitting procedure are shown in Fig.~\ref{fig:suppl:fittingMethodsDisc}~(A-B) for a fixed standoff distance $z = 72\unit{nm}$ (see next paragraph). Estimates for the standard deviations of the fit parameters are obtained from the covariance matrix returned by the fit. 
Experimental data and the corresponding simulations are presented in Fig.~\ref{fig:suppl:fittingMethodsDisc}~(G-J) for a measurement on the $0.6\unit{\um}$ disc.

For the data shown in the main text and in Fig.~\ref{fig:suppl:fittingMethodsDisc}~(A-B), we assume a standoff distance of $z=72\unit{nm}$, based on the fitting results from the channel flow (Fig.~\ref{fig:suppl:fittingMethodsChannel}~(C)). Note that for our discrete set of simulations, the maps computed for $z=72\unit{nm}$ approximate $z_{\mathrm{fit}}\approx 73\unit{nm}$ the best. When additionally fitting for the standoff distance $z$ (Fig.~\ref{fig:suppl:fittingMethodsDisc}~(D-F)), the results are qualitatively similar.  However, we also observe a weak correlation between the fitted standoff $z$ and the disc radius $R$ in this case.  Since all scans were acquired with the same scanning probe, a large change of $z$ is not expected (see also Fig.~\ref{fig:suppl:standoff} for the time evolution of $z$ extracted from the channel fit). Therefore, we believe that this correlation is nonphysical and an artifact of the fitting.

\subsection*{Estimation of $\Dv$, $\lb$ and $z$ from the channel flow}

\begin{figure}[h!]
\includegraphics[width=0.99\textwidth]{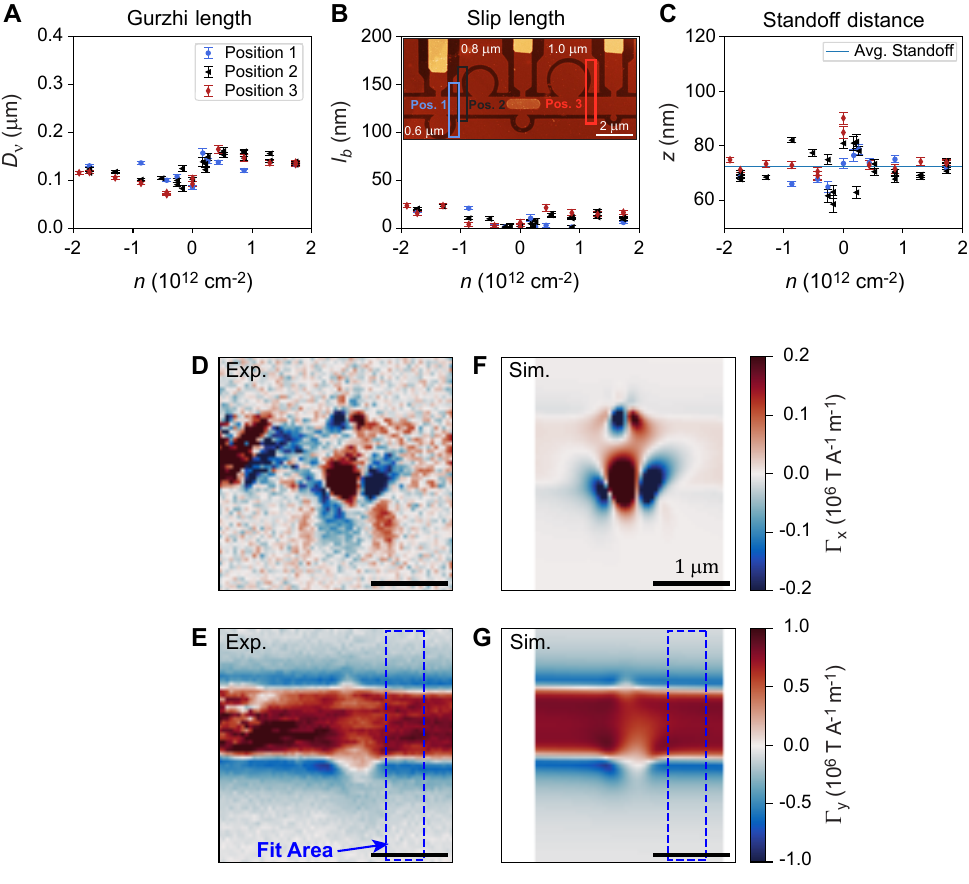}
\caption{\captionstyle{\bf Parameter fits to channel flow.}
	({\bf A-C}) Carrier density dependence of $D_{\nu}$, $l_b$, and $z$ obtained by fitting $\Gamma_y =\frac{1}{I_0} \frac{\Delta B_{\mathrm{NV}}}{\Delta y}$ at three different location above the channel (see inset in {\bf B} for the fit areas). 
	The mean fitted standoff distance is $\approx 73\unit{nm}$. Error bars represent one standard deviation.
	({\bf D} and {\bf E}) Experimental data ($\Gamma_x$, $\Gamma_y$) taken at $n \approx 0.9\ee{12}\unit{cm^{-2}}$ ($\Vbg = 1\unit{V}$) for the $R=0.6\unit{\um}$ disc (same as Fig. \ref{fig:suppl:fittingMethodsDisc}~({\bf G} and {\bf H})). The corresponding simulations with $(D_{\nu}, l_b, z)$ chosen as close as possible to the fitted parameters are shown in {\bf F} and {\bf G}. The fit area is indicated by the blue dashed line.}
\label{fig:suppl:fittingMethodsChannel}
\end{figure}

\begin{figure*}[h!]
	\centering
	\includegraphics[width=0.49\textwidth]{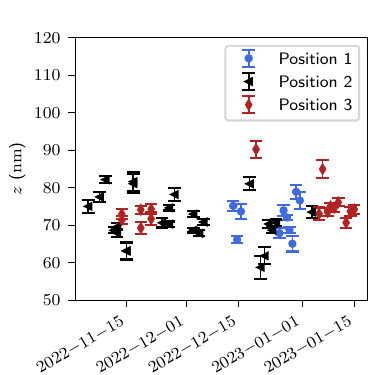}
	\caption{\captionstyle{\bf Standoff distance $z$ of the scans from Fig.~S4 plotted as a function of the measurement date.}}
	\label{fig:suppl:standoff}
\end{figure*}

We further analyze the flow profile through the main channel, which is expected to show a gradual reduction of $J_x$ to zero near the device edges (Poiseuille flow, c.f. supplementary text 2).  Rather than fitting the one-dimensional profile by an analytical function~\cite{ku20}, we apply the above minimization to a two-dimensional channel region, as indicated in Fig.~\ref{fig:suppl:fittingMethodsChannel}.  For the channel fitting, we minimize the magnetic field derivative $\Gamma_y = \frac{1}{I_0}\frac{\Delta \BNV}{\Delta y}$ rather than $\Gamma_x$.
Fig.~\ref{fig:suppl:fittingMethodsChannel}~(A-C) summarizes the results of this analysis.  Interestingly, we find $\approx 2\times$ smaller values for $\Dv$ on the hole side compared to the disc fitting.  Also, the suppression of $\Dv$ near the CNP is much less pronounced.
Entering the parameters obtained from the channel fit of a $R=0.6\unit{\um}$ disc measurement into a simulation of the disc flow, we find that the vortex features are not well reproduced, see \eg~Fig.~\ref{fig:suppl:fittingMethodsChannel}~(D, F).  Thus, a single $\Dv$ value cannot simultaneously and correctly reproduce the vortex and Poiseuille flow profiles.  This points towards a systematic deviation of the observed flow from a purely hydrodynamic (Navier-Stokes) model with rigid boundary conditions.

\subsection*{Systematic error in $\Dv$ due to variations in the device geometry}

\begin{figure}[h!]
\includegraphics[width=0.95\textwidth]{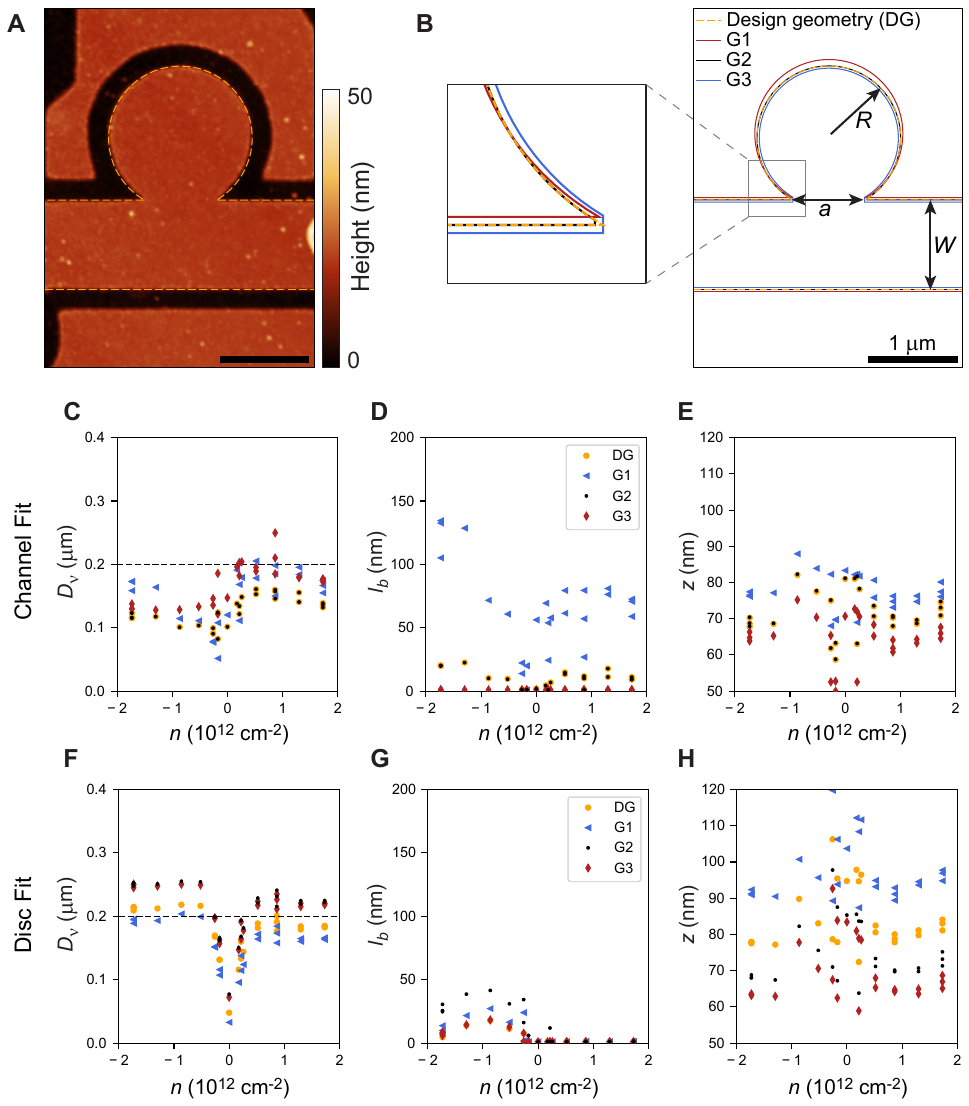}
\caption{\captionstyle{\bf Estimation of the systematic fit errors for the $R=0.8\unit{\um}$ disc.}
	({\bf A}) Height map of the $R=0.8\unit{\um}$ disc acquired with a commercial AFM. The outline of the simulated geometry is indicated by the orange dashed line.
	({\bf B}) Schematic of the different device geometries which are analyzed for estimating the impact of imperfections in the physical device geometry. The orange line represents the geometry as defined in the layout software. G1 represents a slightly larger device with $W=1.05\unit{\um}$ and $R=a=0.825\unit{\um}$. G2 is identical to DG except that the opening gap $a$ is increased by 50 $\unit{nm}$. G3 uses $W=0.95\unit{\um}$, $R=0.775\unit{\um}$ and $a=0.8\unit{\um}$.
	({\bf C-E}) $D_{\nu}$, $l_b$, and $z$ as a function of the carrier density $n$ for the channel data. The fit area is indicated in Fig.~\ref{fig:suppl:fittingMethodsChannel}.
	({\bf F-H}) $D_{\nu}$, $l_b$, and $z$ for the disc data (see Fig.~\ref{fig:suppl:fittingMethodsDisc} for the fit area).
	The dashed lines in {\bf C, F} serve as a guide to the eye.
}
\label{fig:suppl:fittingMethodsSystematic}
\end{figure}

To assess whether this discrepancy could be explained by a mismatch between the simulated and the lithographically-defined device geometry, we generate current density maps for 3 additional device layouts of the $R=0.8\unit{\um}$ disc (see Fig.~\ref{fig:suppl:fittingMethodsSystematic}). As a first test, we study a slightly larger geometry (G1) with $W=1.05\unit{\um}$ and $R=0.825\unit{\um}$ and an opening angle of $60\unit{\deg}$.  This error could be caused, for example, by a slight calibration offset between the commercial AFM used to record the height map that forms the basis for simulations, and the scanning stage of the scanning NV magnetometer.  In a second geometry (G2), we keep the channel width and the disc radius constant but increase the opening gap to $a=0.85\unit{\um}$. Such an modification is expected, for example, if the spatial resolution of the patterning process is insufficient to properly define the sharp corners. Finally, we mimic the case of an overexposure during the e-beam lithography process with geometry G3. For this simulation, we shift the device boundaries inward by $25\unit{nm}$ while keeping disc center at the original location. Note that we keep the opening gap fixed $a=0.8\unit{\um}$ for this study. 

As illustrated in Fig.~\ref{fig:suppl:fittingMethodsSystematic}(C-H), we find that the above variations in the simulated geometry lead to systematic errors in all three fit parameters.  The changes in the extracted $\Dv$ values are approximately $\pm 0.025\unit{\um}$, and exceed the fit errors from the least squares minimization, which are of order $\pm 0.01\unit{\um}$.  Therefore, we conclude that the accuracy of $\Dv$ is dominated by systematic errors related to incomplete knowledge of the device geometry, and not by statistical fit errors.  Since a systematic error shifts all $\Dv$ values in Fig.~4C in the same direction, neither the electron-hole asymmetry nor the pronounced dip near the CNP are affected.  Furthermore, the discrepancies on the hole side between the vortex and channel fits are not eliminated.

\subsection*{Alignment of the device boundary}

To compare experimental with simulated magnetic field maps, we need to accurately determine the physical coordinates of the device with respect to the simulation.  Our two reference coordinates are the y coordinate of the horizontal symmetry axis of the channel ($y_C$) and the x coordinate of the vertical symmetry axis of the circle ($x_C$).  

\begin{figure*}[h!]
\centering
\includegraphics[width=0.99\textwidth]{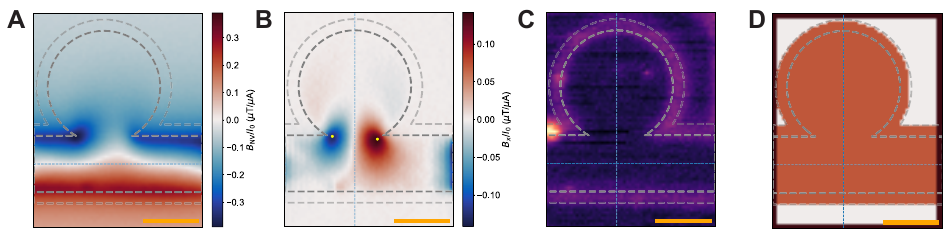}
\caption{\captionstyle{\bf Illustration of the device boundary alignment procedure.}
	($\bf{A}$) $B_{\mathrm{NV}}$ map for determining the horizontal symmetry axis of the channel ($y_C$). 
	($\bf{B}$) $B_{x}$ map for determining the vertical symmetry axis of the circle ($x_C$).  
	($\bf{C}$) NV PL map for validating the boundary alignment. 
	($\bf{D}$) Representation of the mask of the scan used for distinguishing between the scan boundary (black, two pixels), the relevant device region (orange) and the background region (white).  
	The dark (light) gray line corresponds to the boundary of the inner (outer) graphene sheet. The gap is defined via reactive ion etching and has a width of $\sim 0.2\unit{\um}$. 
	Scale bars are $1.0\unit{\um}$}
\label{fig:suppl:methodsAlignment}
\end{figure*}

We have implemented two strategies for determining $y_C$. A first approach (used for Fig.~3, Fig.~\ref{fig:suppl:methodsAlignment}, \ref{fig:suppl:electronholeasymmetries},  and Supplemental Text 5) consists in finding the $y$-coordinates along vertical line cuts where $\BNV$ is closest to zero in the channel. The highest occurrence is then determined to be $y_C$ (see Fig.~\ref{fig:suppl:methodsAlignment}~(A)). A second approach, used for the parameter estimation described in the Methods, consists in finding the maxima and minima of $\BNV$ along vertical line cuts. After fitting the coordinates of the maxima and minima with linear functions, $y_C$ is set to the half-way point between the maximum and minimum locations. For the estimation of $x_C$, we analyze the reconstructed $\Bx$ image (Fig.~\ref{fig:suppl:methodsAlignment}~(B)). The maximum and minimum of the laminar channel flow (marked as yellow dots) should be located symmetrically around the center of the disc. An estimate for $x_C$ is found by averaging the $x$-coordinates of the two extreme values.

The alignment can also be validated by plotting the estimated device boundaries together with the NV photo-luminescence (PL) image recorded simultaneously with the magnetic field map (Fig.~\ref{fig:suppl:methodsAlignment}~(C)). The NV PL map is expected to reflect the device geometry accurately, however, it is less quantitative than the magnetic estimation above. The physical device edges indicated in Fig~4A, Fig.~\ref{fig:suppl:CNPRamsey}, \ref{fig:suppl:vortexLinecombined}, and \ref{fig:suppl:EField} are determined directly from the PL maps.

\subsection*{Current monitoring normalization}

\begin{figure*}[h!]
\centering
\includegraphics[width=0.79\textwidth]{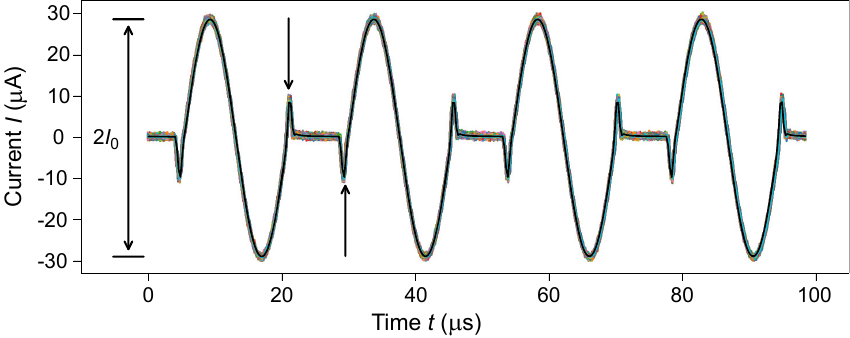}
\caption{\captionstyle{\bf Extraction of the current amplitude from experimental data.}
	The current signal $I(t)$ is recorded at each pixel (colored traces). The current amplitude $I_0$, defined as half of the peak-to-peak current signal, is then extracted from the average over all traces (black). The sharp peaks (indicated by arrows) are due to a modulation of the back-gate voltage. Since they occur outside the phase accumulation window of the quantum sensor, they do not influence the magnetometry signal. Four repetitions are shown, corresponding to the four readout phases of the sensing protocol.}
\label{fig:suppl:methodsCurrent}
\end{figure*}

We monitor the device current by recording a sample $I(t)$ of the source-drain current at each pixel, see Fig.~\ref{fig:suppl:methodsCurrent}.  The amplitude $I_0$ is determined as one-half the peak-to-peak amplitude of $I(t)$.  When comparing current flow patterns, we typically normalize the magnetic field maps by $I_0$.

\subsection*{Measurement parameters for Figs. 2-4}

Fig.~2A used the following parameters: 
$n\approx-1.7\ee{12}\unit{cm^{-2}}$ ($\Vbg = -2\unit{V}$), 
$I_0 = 28.7\unit{\uA}$. $\theta=56^\circ$, $\varphi=1^\circ$ (Scanning probe NV1). For the streamlines, we reconstruct the current density with $z = 75\unit{nm}$ and $\lambda = 1.5\cdot z$.

Fig.~3A used the following experimental parameters: 
$n\approx-1.7\ee{12}\unit{cm^{-2}}$ ($\Vbg = -2\unit{V}$), 
$I_0 = \{15.7, 15.7, 15.5, 15.4 \} \unit{\uA}$ from left to right. $\theta=55^\circ$, $\varphi=1^\circ$ (Scanning probe NV2). 

Fig.~3B used the following simulation parameters: $D_\nu = 0.28\unit{\um}$, $z = 110\unit{nm}$. 

Fig.~4A used the following parameters: 
$n\approx \{-0.9, 0, +0.9\} \cdot 10^{12}\unit{cm^{-2}}$ ($\Vbg = \{-1,0,+1\}\unit{V}$) from left to right, 
$I_0 = \{24.3, 8.7, 24.3\} \unit{\uA}$ from left to right. $\theta=56^\circ$, $\varphi=1^\circ$ (Scanning probe NV1).

Fig.~4C used $n\approx (-1.9\dots 1.7) \cdot 10^{12}\unit{cm^{-2}}$ corresponding to $\Vbg = (-2.2 \dots +2)\unit{V}$, 
$I_0 = 2-15 \unit{\uA}$. All measurements were acquired with scanning probe NV2. We fit the experimental data assuming $z = 72\unit{nm}$. 

\clearpage
\section*{Supplementary Text 1: Non-linearity near charge neutrality}

\begin{figure}[h!]
\centering
\includegraphics[width=0.89\textwidth]{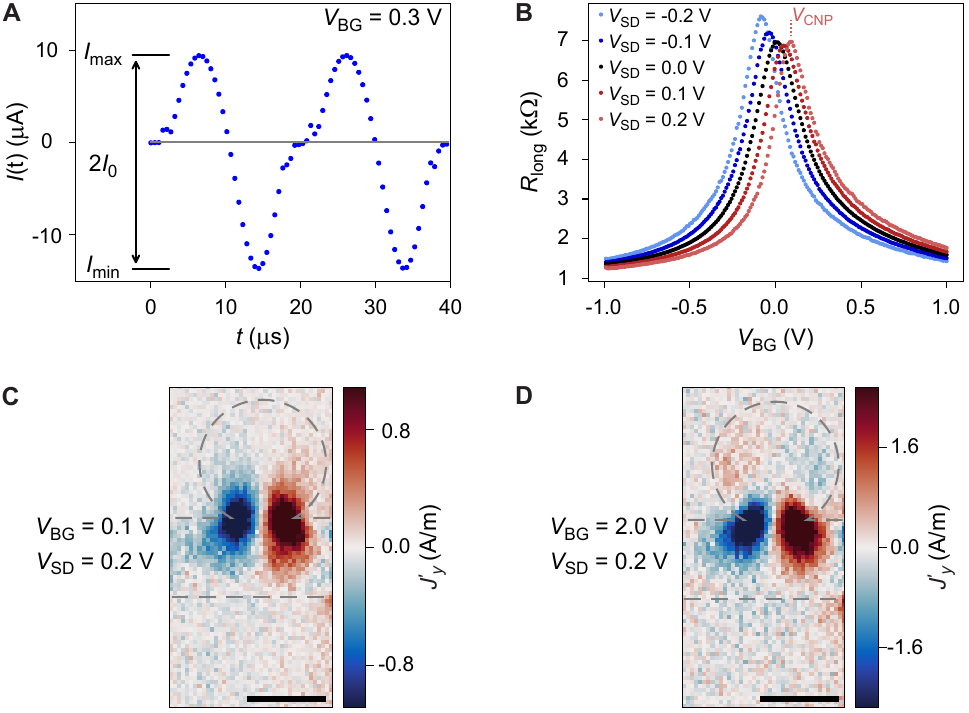}
\caption{\captionstyle{\bf Imaging near charge neutrality.}
	({\bf A}) Asymmetry in the current trace of a scan recorded at $V_{\mathrm{BG}} = 0.3\unit{V}$ and $V_{\mathrm{SD}} = 0.2\unit{V}$. 
	({\bf B}) Longitudinal resistance measured between contacts C and D as a function of the back-gate voltage for different source-drain biases. A small AC modulation ($V_{\mr{AC}} = 5\unit{mV}$) is added on top of the bias voltage for lock-in detection.
	({\bf C}) DC map of $\JyTilde$ acquired close to the CNP for a source-drain voltage of $V_{\mathrm{SD}} = 0.2\unit{V}$. 
	({\bf D}) Map of $\JyTilde$ acquired away from charge neutrality. For this image, the averaging time per pixel has been reduced to yield approximately the same SNR as the image shown in ({\bf C}). 
	Scale bars are $1\unit{\um}$.}
\label{fig:suppl:CNPRamsey}
\end{figure}

In the vicinity of the CNP, we often observe an asymmetry in the recorded current trace for a symmetrically applied AC source-drain voltage (Fig.~\ref{fig:suppl:CNPRamsey}~(A)). This could be an indication that the local carrier density is not fixed during the AC magnetometry protocol.
Since we apply source-drain voltages on the order of $100\unit{mV}$ to overcome the large two-terminal resistance at the CNP ($\sim 25\unit{kOhm}$) and generate a detectable source-drain current, the electrostatic potential at the measurement location is also expected to change.  Indeed, measurements of the longitudinal resistance between contacts C and D (comprising the Au patch, Fig.~\ref{fig:suppl:fab}) confirm that the local $V_{\mathrm{CNP}}$ changes as a function of the applied bias voltage (Fig.~\ref{fig:suppl:CNPRamsey}~(B)).

To exclude that the fading of the vortex feature close to charge neutrality is an artifact of a carrier density modulation, we image the $R=0.8\unit{\um}$ disc using a complementary DC technique. For this purpose, we conduct a Ramsey-type experiment at $V_{\mathrm{SD}} = 0.2\unit{V}$ with $\tau = 16 \cdot 2\pi/A_g^{||} \approx 5.27\unit{\us}$. Here, $A_g^{||} \approx 2\pi \cdot 3.03\unit{MHz}$ is the parallel hyperfine coupling of the NV center. For this particular choice of the evolution time $\tau$, the polarization of the $^{15}$N nuclear spin forming the NV center does not affect the measurement result and a simple PL signal of the form of Eq.~\ref{eq:theory:phase} is recovered. We take measurements near charge neutrality ($V_{\mathrm{BG}} = 0.1\unit{V}$) and far away ($V_{\mathrm{BG}} = 2\unit{V}$) using a differential scheme (signal on/off). The resulting $\JyTilde$ maps are shown in Fig.~\ref{fig:suppl:CNPRamsey}~(C-D) for a DC current flowing in negative $x$ direction. This experiment confirms that the vortex feature indeed disappears near charge neutrality.

\section*{Supplementary Text 2: Poiseuille flow}

We also analyze the channel flow profile, which should turn from rectangular to parabolic as the transport changes from diffusive to hydrodynamic.  This spatial signature is known as Poiseuille flow, and has been analyzed in previous spatial imaging experiments~\cite{sulpizio19,ku20}. In the limit where the current density vanishes completely at the device boundaries (no-slip), the current profile is given by \cite{torre15}:
\begin{equation}
	\Jx = \frac{\sigma_0}{e}\nabla \phi \left[ 1 - \frac{\cosh{\frac{y-y_0}{D_{\nu}}}}{\cosh{\frac{w}{2D_{\nu}}}}\right]
\end{equation}
In the extreme case where $D_{\nu}$ is large compared to the width $w$ of the channel (and $l_{\mathrm{ee}} \ll w$), the current profile can be described approximately by a parabola.  Ref.~\cite{ku20} observed such behavior in room-temperature monolayer graphene and reported $D_{\nu}\gtrsim 0.3\unit{\um}$ for their devices. Given the estimate for the Gurzhi length in our device (see Figs.~\ref{fig:suppl:fittingMethodsDisc}, \ref{fig:suppl:fittingMethodsChannel}), we would expect to observe a non-uniform channel profile at the very least away from the CNP, \ie, for $|n| \gtrsim 0.5\ee{12}\unit{cm^{-2}}$.

\begin{figure}[h!]
\includegraphics[width=0.99\textwidth]{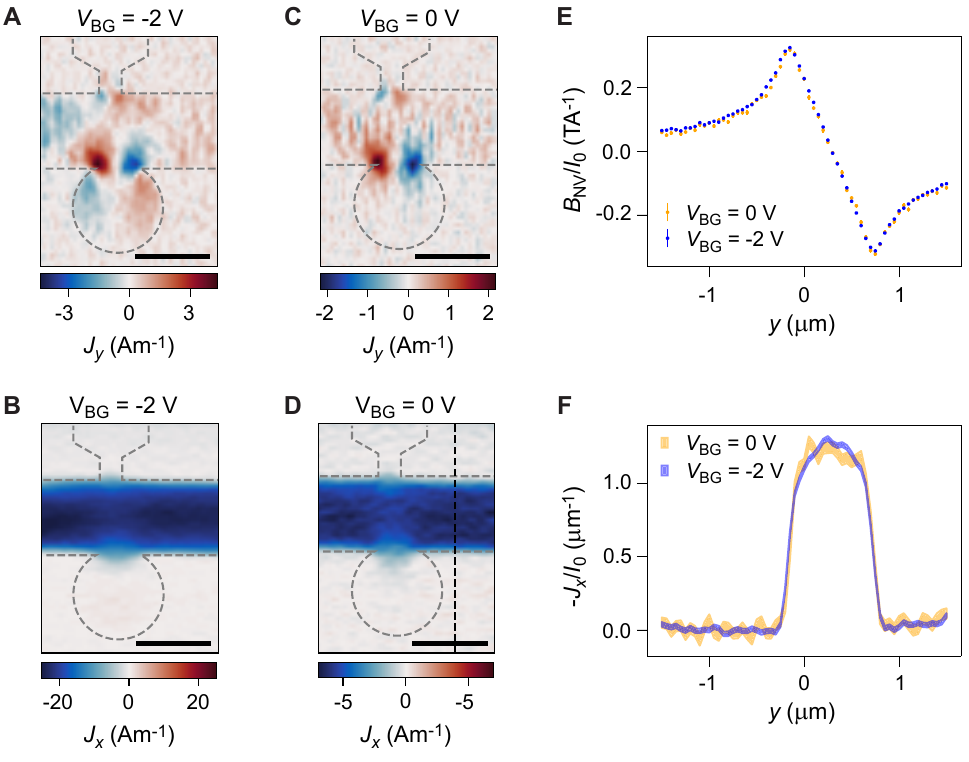}
\caption{\captionstyle{\bf Simultaneous DC imaging of current vortices and channel profiles}. 
	({\bf A} and {\bf B}) Maps of the current density components $\Jx$ and $\Jy$ at $V_{\mathrm{BG}} = -2\unit{V}$ ($n \approx -1.7\ee{12}\unit{cm^{-2}}$).  Measurements use a Ramsey protocol with a phase accumulation time of $\tau = 11.8\unit{\us}$.
 	For the reconstruction, we assume a standoff distance of $z = 72\unit{nm}$ (estimated based on the fitting results for the channel) and use $\lambda = 1.5\cdot z$. The device geometry is indicated with dashed lines (estimated from PL maps).
	({\bf C} and {\bf D}) Corresponding maps for $V_{\mathrm{BG}} = 0\unit{V}$. The black line indicates the location of the line cuts analyzed in {\bf E} and {\bf F}. Scale bars are $1\unit{\um}$.
	({\bf E}) Comparison of the magnetic field line scans across the channel. The data sets are normalized by the device current $I_0$. Error bars represent one standard deviation, extracted from the shot noise in the measurement signal \cite{palm22}.
	({\bf F}) Reconstructed current density profiles. The shaded area represents one standard deviation. This uncertainty is extracted from a region without signal next to the channel.}
\label{fig:suppl:vortexLinecombined}
\end{figure}

Fig.~\ref{fig:suppl:vortexLinecombined} shows maps of the current density in the $R=0.6\unit{\um}$ disc at $V_{\mathrm{BG}} = -2\unit{V}$ ($n \approx -1.7\ee{12}\unit{cm^{-2}}$) (A-B), and at $V_{\mathrm{BG}} = 0\unit{V}$ (C-D). These maps were recorded using the Ramsey protocol to prevent a modulation of the carrier density during data acquisition (see Supplementary Text~1). Again, we observe a current vortex only away from charge neutrality. Line cuts of the normalized magnetic field $B_{\mathrm{NV}}/I_0$ and current density $\Jx/I_0$ are shown in Fig.~\ref{fig:suppl:vortexLinecombined}~(E-F). We notice that $B_{\mathrm{NV}}$ barely differs between the two images. Fortunately, the reconstructed current density is more instructive. The channel profile at hole doping is indeed more parabolic than at charge neutrality. While this observation is consistent with our previous findings, the differences are less striking than the presence or absence of a current vortex. This is to be expected, because the channel profile becomes gradually flatter in the center upon decreasing $D_{\nu}$ and does not display a hallmark sign change like the vortex maps. Therefore, we find the whirlpools to be better suited for studying electron hydrodynamics in our device than the Poiseuille flow.

\section*{Supplementary Text 3: Electric field imaging}

\begin{figure}[h!]
\includegraphics[width=0.9\textwidth]{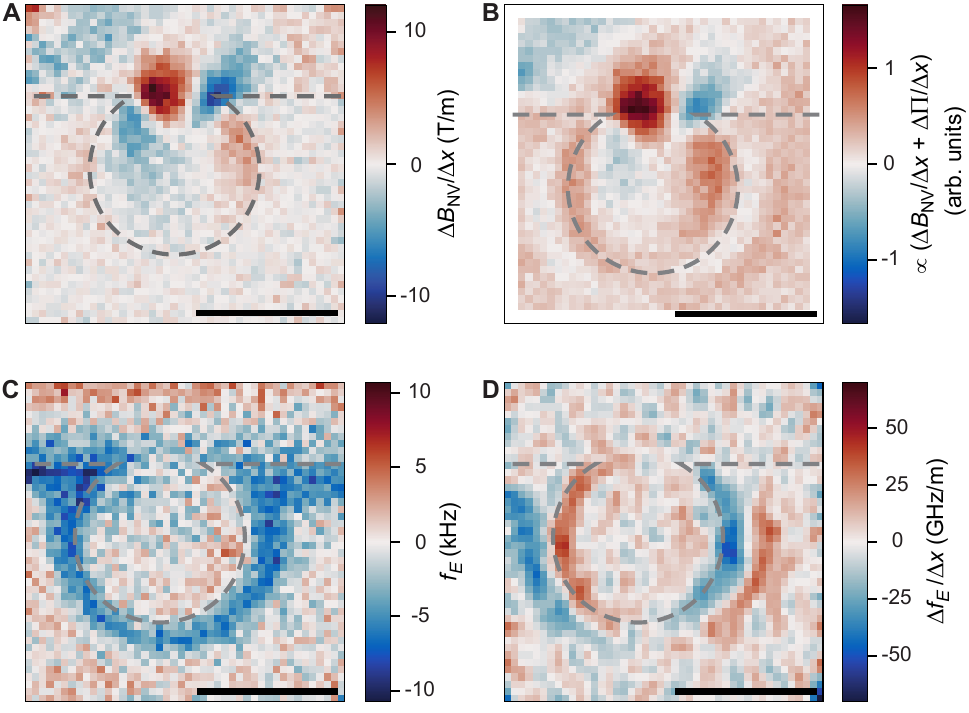}
\caption{\captionstyle{\bf Imaging of the magnetic and electric field over the $R=0.6\unit{\um}$ disc.}
	({\bf A}) Map of $\Delta B_{\mathrm{NV}}/\Delta x$ derived from an AC measurement at $n \approx -1.7\ee{12}\unit{cm^{-2}}$.
	({\bf B}) Image of the same region recorded using the gradiometry technique. An additional gradient signal $\Delta \Pi / \Delta x$ is picked up, most prominently at the device edge.
	({\bf C}) Map of the shift $f_E$ of the NV resonance frequency caused by the electric field from the back gate. An AC detection scheme is used for this measurement.
	({\bf D}) Spatial derivative along $x$ of the map shown in ({\bf C}). This map is low-pass filtered in Fourier space using a Hann filter with a cutoff frequency at $2\pi / (100\unit{nm})$. 
	The dashed lines indicate the physical edge of the device and have been determined from PL maps. Scale bars are $1\unit{\um}$.}
\label{fig:suppl:EField}
\end{figure}

Fig.~\ref{fig:suppl:EField}~(A-B) displays the numerically computed derivative of a measured AC magnetic field map acquired using the protocol described in the Methods section, together with a corresponding image obtained using the scanning gradiometry technique~\cite{huxter22}.  Both measurements are taken at $n \approx -1.7\ee{12}\unit{cm^{-2}}$.  As expected, the vortex appears in both images.  However, the gradiometry technique picks up an additional signal above the etched region of the vdW stack where the back gate is not screened by the graphene sheet.  We attribute this signal to the static electric field generated by the back-gate potential.  Such static electric fields are only visible in a dynamic imaging mode (oscillating tuning fork) because they are otherwise screened by mobile charges on the diamond tip~\cite{huxter23, qiu22}.

To confirm the electrical origin of this signal, we image the sample again using an AC sensing technique (Hahn echo). However, instead of modulating the device current, we modulate the back-gate voltage $\Vbg$. A map of the resulting electric-field-induced frequency shift and its derivative along $x$ are shown in Fig.~\ref{fig:suppl:EField}~(C-D). These maps clearly show the presence of an electric field above the etched part of the device. Furthermore, the features observed in the gradiometry scan are qualitatively well explained by the electric field gradient.

\section*{Supplementary Text 4: Effect of an out-of-plane magnetic field}
\label{sec:external_field}

Current flow in the hydrodynamic model, subject to an out-of-plane magnetic field $B_z$, is described by the linearized Navier-Stokes equation and the continuity equation \cite{pellegrino17, berdyugin19}:
\begin{align}
	\vec{J}(\vec{r}) - D_{\nu}^2\nabla^2\vec{J}(\vec{r}) + \omega_c \tau  (1+D_{\mathrm{H}}^2&\nabla^2)\vec{J}(\vec{r})\times\vec{e_z}  + \sigma_0\nabla\phi(\vec{r}) = 0 \\
	\nabla\cdot \vec{J}(\vec{r}&) = 0
	\label{appendixHydro:NavierStokes}
\end{align}
In this equation, $\omega_c = \mathrm{sgn}(n)\frac{e B_z}{m^*}$ is the cyclotron frequency, $\tau$ is the mean free time with respect to momentum-relaxing scattering events, and $m^*$ is the cyclotron mass.  We include the signum function $\mathrm{sgn}(n)$ to reproduce the correct sign dependence for electrons ($n > 0$) and holes ($n < 0$). $D_{\mathrm{H}}$ is a diffusion constant related to the Hall viscosity $\nu_{\mathrm{H}}$ \cite{pellegrino17,  berdyugin19}.

\begin{figure}[h!]
\includegraphics[width=0.8\textwidth]{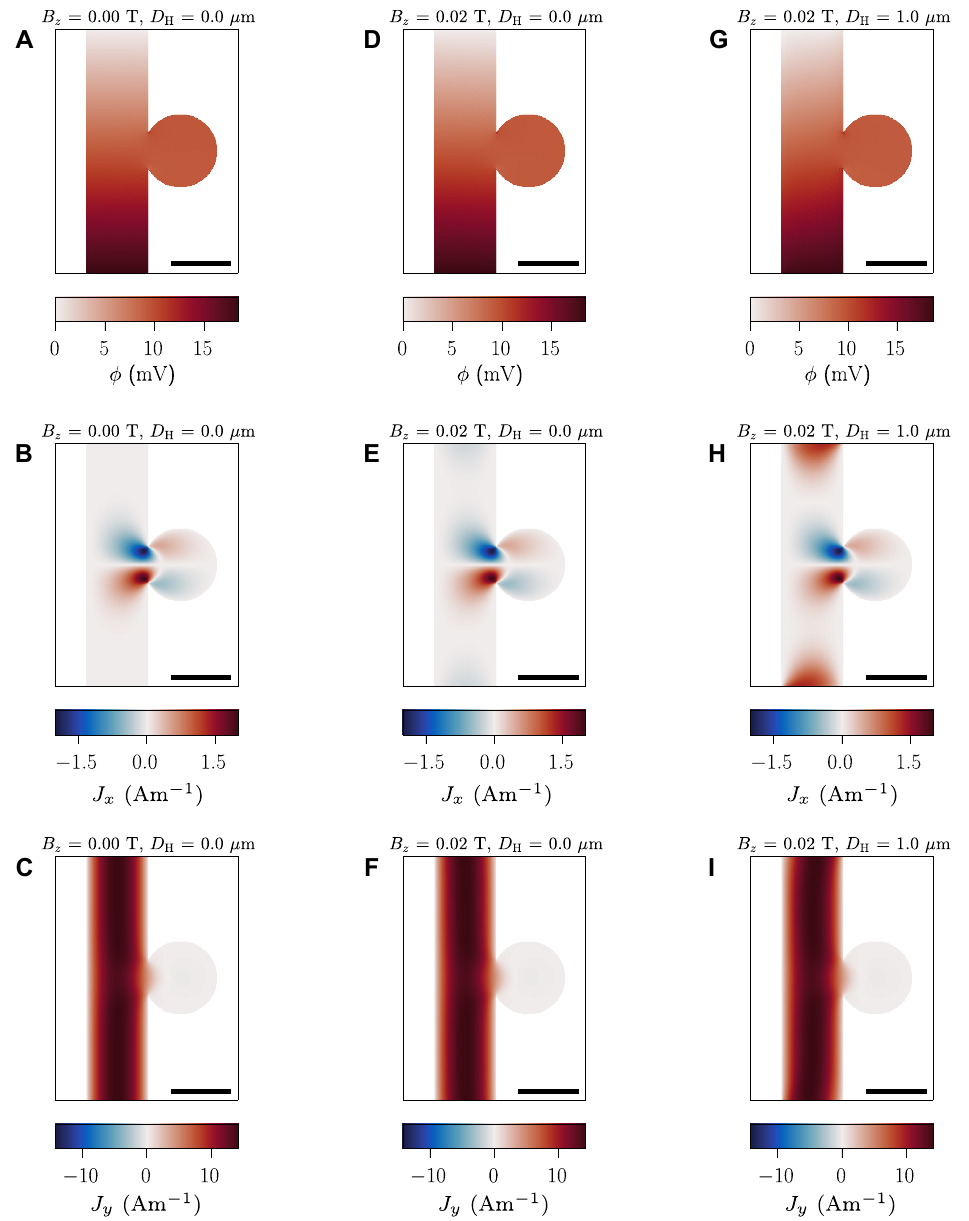}
\caption{\captionstyle{\bf Simulations of the electric potential $\phi$ and current density components $\Jx$, $\Jy$ in a perpendicular magnetic field.}
	({\bf A}-{\bf C}) Solutions of the Navier-Stokes equation in zero-field.
	({\bf D}-{\bf F}) Simulation results for an out-of-plane magnetic field of $B_z = 20\unit{mT}$. 
	({\bf G}-{\bf I}) Simulation results at $B_z = 20\unit{mT}$ and assuming $D_{\mathrm{H}} = 1\unit{\um}$. 
	For all simulations, we assume $n = 10^{12}\unit{cm^{-2}}$, $D_{\nu} = 0.25\unit{\um}$ and $\mu = e\tau/m^* = 2.64\cdot10^4\unit{cm^2/Vs}$.}
\label{fig:suppl:hydroBSim}
\end{figure}

In a typical scanning NV magnetometry experiment, applied magnetic fields do not exceed a few tens of $\unit{mT}$ and expected values for the diffusion length $D_{\mathrm{H}}$ are $< 1\unit{\um}$ for monolayer graphene at room temperature \cite{berdyugin19}. While a perpendicular magnetic field does affect the potential landscape, it does not significantly change the current profile in the Hall-bar geometry. As shown in Fig.~\ref{fig:suppl:hydroBSim}, the current density distribution is only modified near the source and drain contacts, but not in the imaging region near the discs.  Therefore, we neglect the pertinent terms in the equations of the main text and the associated simulations. 

\section*{Supplementary Text 5: Relativistic lattice Boltzmann simulations}

\begin{figure*}[h!]
\centering
\includegraphics[width=0.8\textwidth]{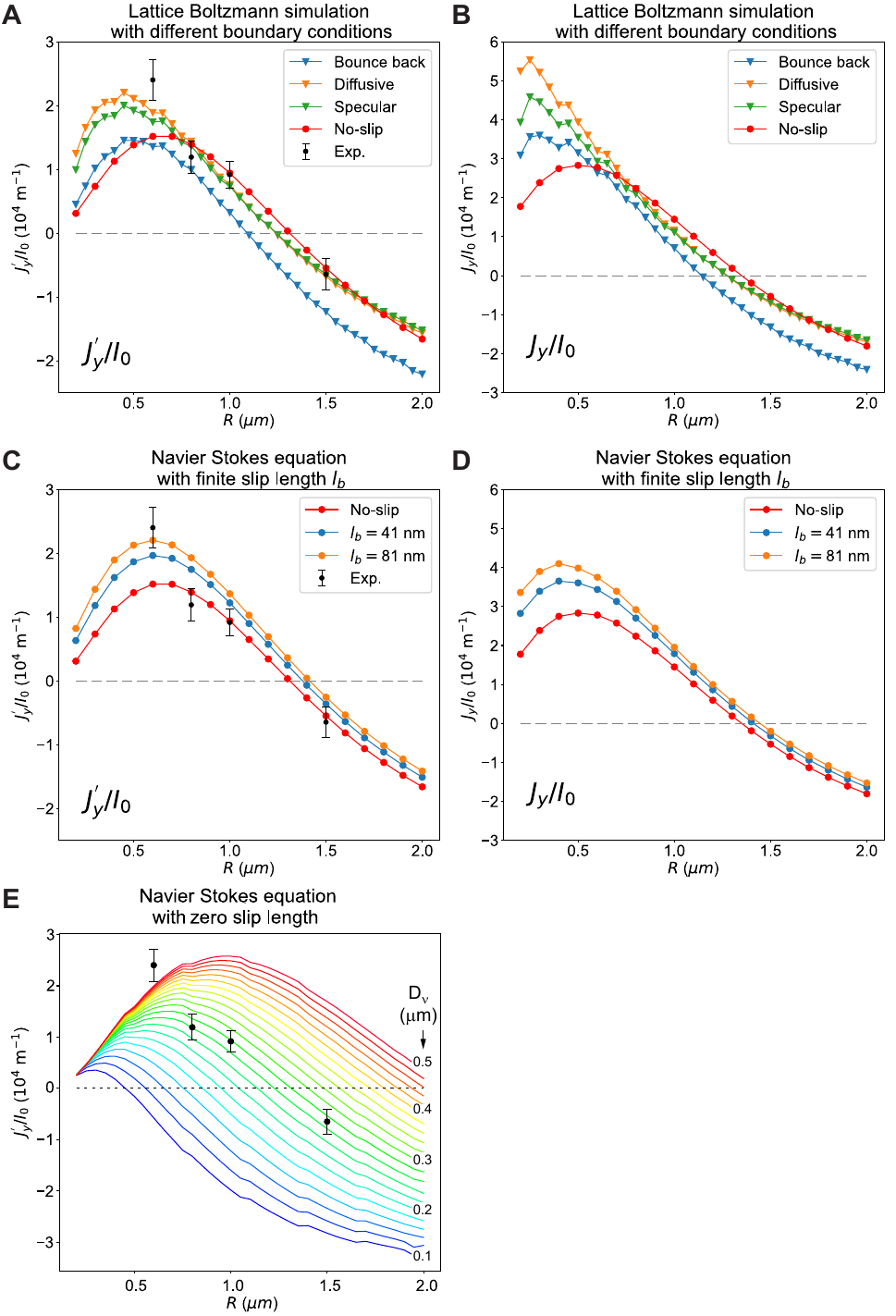}
\caption{\captionstyle{\bf Disc simulations beyond the no-slip Navier-Stokes simulation and for the Lattice-Boltzmann method.}
	({\bf A} and {\bf B}) Normalized low-pass filtered current density $\JyTilde/I_0$ ({\bf A}) and true current density $\Jy/I_0$ ({\bf B}) obtained from lattice Boltzmann simulations with different boundary conditions.
	({\bf C} and {\bf D}) Normalized low-pass filtered current density $\JyTilde/I_0$ ({\bf C}) and true current density $\Jy/I_0$ ({\bf D}) obtained from Navier-Stokes simulations with different slip lengths.
	In all panels, the no-slip data (shown with red circles) is the Navier-Stokes simulation curve shown in Fig.~3E of the main text and the black dots are the corresponding data points.
	({\bf E}) $\JyTilde/I_0$ for a no-slip boundary condition and varying $D_{\nu}$.}
\label{fig:suppl:LBM}
\end{figure*}

\begin{figure*}[h!]
\centering
\includegraphics[width=0.99\textwidth]{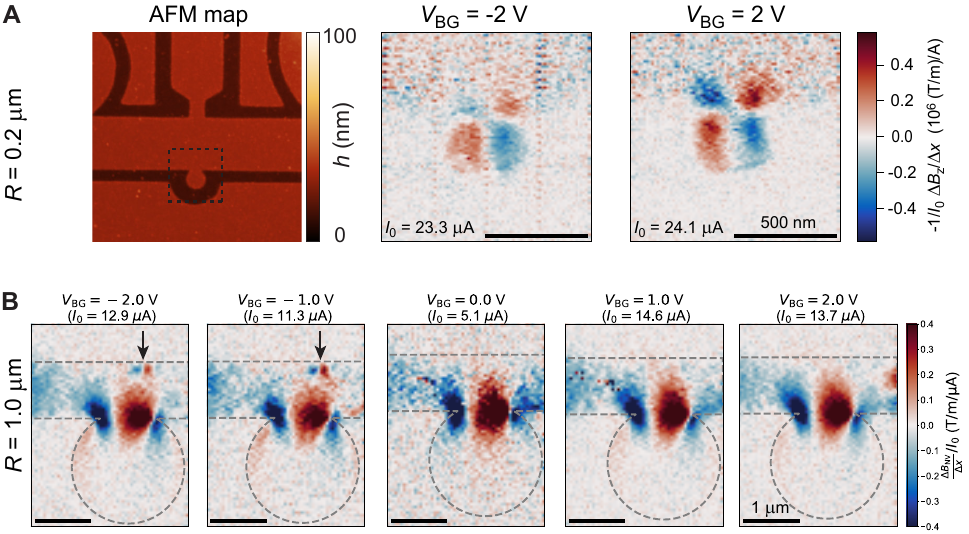}
\caption{\captionstyle{\bf Observed asymmetries between electron and hole doping.}
	({\bf A}) Asymmetry in the vortex flow for the $0.2\unit{\um}$ disc. The signature in the channel is much less pronounced for the scan at $\Vbg=-2\unit{V}$ ($n \approx -1.7\ee{12}\unit{cm^{-2}}$). For these measurements, we use a dynamic decoupling sequence with 8 refocusing pulses and a phase accumulation time of $\tau = 55\unit{\us}$.
	({\bf B}) Carrier-type dependent scattering at the device edge for scans on the $R=1.0\unit{\um}$ disc. The arrow indicates the location where the asymmetry occurs.}
\label{fig:suppl:electronholeasymmetries}
\end{figure*}

We employ the relativistic lattice Boltzmann method (RLBM) to model two-dimensional (2D) single-particle flow away from the hydrodynamic and diffusive limits, shown with Fig.~\ref{fig:suppl:LBM}. Our RLBM framework is based on the D2V72 quadrature scheme~\cite{coelho18rel} in the ultra-relativistic limit~\cite{anderson74}. RLBM simulations discretize the energy and momentum phase space of the quasi-particle distribution function $f$ into a set of 72 quadrature components. These quadratures are combinations of six energies and twelve isotropic momenta vectors in 2D. $f$ is uniquely defined at every lattice point and lattice points are placed on a square grid to approximate any simulation geometry in real space. The governing equation is given by:
\begin{equation} \label{eq:rlbm_tensor}
	p^\mu \partial_\mu f = \frac{p_\mu U^\mu}{\vF^2}\Omega[f]~,
\end{equation}
where $p^\mu = |p| [1, \frac{v_x}{\vF},\frac{v_y}{\vF}]$ is the quasi-particle momentum, $U^\mu = \gamma[\vF, u_x, u_y]$ is the macroscopic velocity, $\gamma = \left(1 - {\bf u}\cdot {\bf u}/\vF^2 \right)^{-1/2}$ is the Lorentz factor, and $\partial^\mu = [\frac{\partial_t}{\vF}, -\partial_x, -\partial_y]$ is the gradient operator, shown in contravariant form using the $(+,-,-)$ metric signature. $\bf{u} = [u_x, u_y]$ and $\bf{v}  = [v_x, v_y]$ are the macroscopic (belonging to the lattice point) and microscopic (belonging to the individual quadrature) two-component velocities, respectively. $\vF \sim 10^6\unit{m/s}$ is the Fermi velocity (analogous to the speed of light in special relativity), and $\Omega$ is the collision operator, defined below. It is convenient to convert Eq.~\ref{eq:rlbm_tensor} into the following version of the RLBM equation:
\begin{equation} \label{eq:rlbm}
	\frac{\partial f}{\partial t} + {\bf v} \cdot \nabla f = \eta \Omega[f]~,
\end{equation}
as it closely resembles the classical LBM equation with one addition prefactor term, $\eta = \gamma\left( 1 - {\bf v}\cdot {\bf u}/\vF^2\right)$, that captures relativistic effects. The left- and right-hand sides of Eq.~\ref{eq:rlbm} represent the streaming and collision steps of the simulation, respectively. In the streaming step, the quadratures are propagated outward from their respective lattice point according to their momenta and are collected by neighboring lattice points. Bilinear interpolation is used to collect streamed quadrature components that end up between neighboring lattice points~\cite{coelho18rel}. In the collision step, the collected quadrature components are redistributed by the collision operator according to their energy and momentum. The RLBM framework alternates between streaming and collision steps to iteratively approach a steady-state distribution across all lattice points.

The collision operator includes both momentum-conserving and relaxing terms to account for the carrier-carrier scattering and carrier-phonon/impurity scattering. Specifically, we set
\begin{equation}
	\Omega[f_k] = \frac{f_k^{ee} - f_k}{\tee} + \frac{f_k^{mr} - f_k}{\tmr} = \frac{f_k^{ee}}{\tee} + \frac{f_k^{mr}}{\tmr} -\frac{f_k}{\tau_\mathrm{eff}}~,
\end{equation}
where the subscript $k$ refers to the 72 quadrature components, $f_k^{ee}$ comes from evaluating the Fermi-Dirac equilibrium distribution function (using the BGK approximation~\cite{bhatnagar54}) and $f_k^{mr}$ isotropically redistributes the momentum at every lattice point via energy-conserving collisions~\cite{kashuba18}. It can be explicitly written as
\begin{equation}
	f_k^{mr} = \frac{\sum_i \delta_{\varepsilon_k,\varepsilon_i} f_i}{\sum_i \delta_{\varepsilon_k,\varepsilon_i}}~,
\end{equation}
where $\delta$ is the Kronecker delta function and $\varepsilon_k$ is the energy of the $k^\mathrm{th}$ quadrature. The $f_k/\tau_\mathrm{eff}$ term ensures quasi-particle conservation.
The simulation-wide time constants directly relate to the macroscopic scattering length scales through $\tee = l_{ee}/\vF$, $\tmr = l_{mr}/\vF$, with $\tau_\mathrm{eff}^{-1} = \tee^{-1} + \tmr^{-1}$. Thus, $\tee$ and $\tmr$  act as user-controlled values that steer the RLBM simulation towards a more hydrodynamic or diffusive behavior. Ballistic effects, while not directly accounted for, naturally arise as characteristic device sizes decrease relative to all scattering lengths. We note that simulation artifacts may arise if ballistic effects dominate (\eg, in the deep ballistic regime).

We describe scattering off device edges via one of three redistribution methods. Bounce-back scattering~\cite{kruger16}, where momentum is inverted, is used to mimic a zero-slip-length boundary condition as it ensures zero velocity on the edges. Specular scattering~\cite{dejong95}, which reflects perpendicular components of momenta, and diffusive scattering~\cite{ansumali02}, which redistributes momenta accounting for energy and density conservation, mimic a more general boundary condition with non-zero slip length.

To set the device current, we apply Neumann boundary conditions (a constant, uniform flux of current density) at the source and drain contact of the simulated device. No additional forcing term was applied. A given simulation iterates until subsequent iterations show an average absolute change in macroscopic velocity across all lattice points that is below a convergence threshold, typically of order $10^{-6}$. Other simulation details include a lattice grid size of $25\unit{nm}$, $T = 300\unit{K}$ and no chemical doping.

To determine the macroscopic current density, we first compute the energy-momentum tensor $T^{\mu\nu}$ at every lattice point~\cite{coelho18rel},
\begin{equation}
	T^{\mu\nu} = \sum_k f_k p_k^\mu p_k^\nu~.
\end{equation}
Then, we solve the eigenequation $T^{\mu}_\nu U^{\nu} = \varepsilon U^{\mu}$ numerically (via the power method) for the macroscopic energy density eigenvalue $\varepsilon$ and the macroscopic velocity eigenvector $U^{\mu}$. From the macroscopic velocity, the quasi-particle density can then be computed with $\rho = \sum_k U_\mu p_k^\mu f_k =  \vF  \sum_k f_k |p_k|\eta_k$. Finally the current density ${\bf J}$ can be obtained by combining the charge $q$, density $\rho$, and macroscopic two-component velocity $\bf{u}$:
\begin{equation}
	{\bf J} = q \rho {\bf u} = q \vF \sum_k f_k |p_k|\eta_k {\bf u}~.
\end{equation}


\clearpage
\section*{Supplementary Figure 1}

\clearpage
\section*{Supplementary Figure 2}

\clearpage
\section*{Supplementary Figure 3}

\clearpage
\section*{Supplementary Figure 4}

\clearpage
\section*{Supplementary Figure 5}

\clearpage
\section*{Supplementary Figure 6}

\clearpage
\section*{Supplementary Figure 7}

\clearpage
\section*{Supplementary Figure 8}

\clearpage
\section*{Supplementary Figure 9}

\clearpage
\section*{Supplementary Figure 10}

\clearpage
\section*{Supplementary Figure 11}

\clearpage
\section*{Supplementary Figure 12}

\clearpage
\section*{Supplementary Figure 13}

\clearpage
\section*{Supplementary Figure 14}

\clearpage
\section*{Supplementary Figure 15}
\begin{figure*}[h!]
\centering
\includegraphics[width=0.9\textwidth]{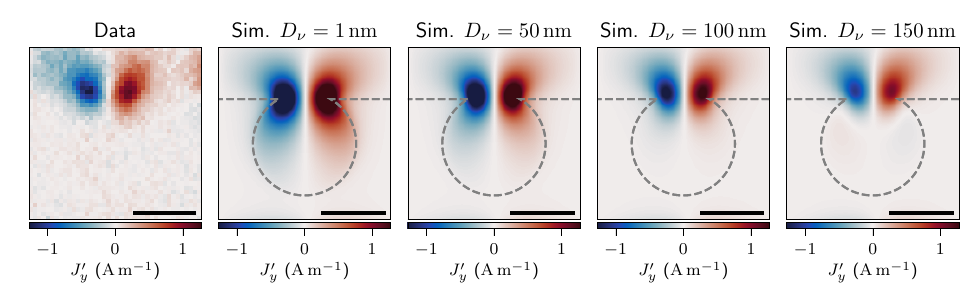}
\caption{\captionstyle{\bf Demonstration that the Gurzhi length is non-zero even at charge neutrality.}
	Plots show the measured data ($\Vbg = 0\unit{V}$) together with the corresponding simulations of the Navier-Stokes equation assuming different values for $D_{\nu}$. A no-slip boundary condition is assumed for all simulations.}
\label{fig:suppl:CNPVsSims}
\end{figure*}

\newpage
\section*{References}